\documentclass{aa}  
\usepackage{aalongtable}
\usepackage{graphicx}
\usepackage{amsmath}
\usepackage{txfonts}
\usepackage[colorlinks=true,
            urlcolor=blue,
            linkcolor=blue,
            citecolor=blue]
            {hyperref}
\usepackage{threeparttable}

\newcommand{\swift}{\textit{Swift}}
\newcommand{\xmm}{XMM-\textit{Newton}}
\newcommand{\nustar}{NuSTAR}
\newcommand{\iras}{{IRAS\,23226-3843}}
\newcommand{\plm}{$\pm$}
\newcommand{\rb}[1]{\raisebox{1.5ex}[-1.5ex]{#1}}
\newcommand{\Ha}{H$\alpha$}
\newcommand{\Hb}{H$\beta$}

\newcommand{\mcc}[1]{\multicolumn{1}{c}{#1}}

\newcommand{\kms}{km\,s$^{-1}$}

\begin{document} 

    \title{Extreme line profile variations in the repeating changing-look active galactic nucleus \iras{}}

   \author{W. Kollatschny \inst{1}, 
           D. Grupe \inst{2},
           H. Winkler \inst{3},
           M. A. Probst \inst{1},        
           M. W. Ochmann \inst{1},
           A. P\"onitzsch \inst{1}, 
           N. Schartel \inst{4},
           S. Wolsing \inst{1,2}, 
           S. Komossa \inst{5},
           S.B. Potter \inst{3,6}
         }

   \institute{Institut f\"ur Astrophysik und Geophysik, Universit\"at G\"ottingen,
              Friedrich-Hund-Platz 1, 37077 G\"ottingen, Germany\\
              \email{wkollat@gwdg.de}
         \and
         Department of Physics, Geology, and Engineering Technology, Northern Kentucky University, Highland Heights, KY 41076, USA
          \and
         Department of Physics, University of Johannesburg, PO Box 524, 2006 Auckland Park,
         Johannesburg, South Africa    
         \and
         ESA, European Space Astronomy Centre (ESAC), 28692
         Villanueva de la Ca{\~nada}, Madrid, Spain
          \and
         Max-Planck-Insitut f\"ur Radioastronomie, Auf dem H\"ugel 69,
         D-53121 Bonn, Germany
          \and
        South African Astronomical Observatory, Observatory Road, Observatory, 7925, Cape Town, RSA
          }

   \date{Received 13 November 2025; Accepted 18 January 2026 }
   \authorrunning{Kollatschny et al.}
   \titlerunning{The changing-look AGN \iras}

 
  \abstract
  {}
   {\iras{} has been identified as a highly variable Seyfert galaxy 
   and even as a changing-look active galactic nucleus
   based on optical spectra. 
   Here we present follow-up observations -- taken over the past five years -- for examining the ongoing photometric and spectral variations in this remarkable galaxy.}
   {We carried out \swift{} observations of \iras{} together 
   with new optical spectra taken in 2023 and 2024. In parallel we investigate ASAS-SN photometric data from 2014 till 2025.} 
   {\iras{} stayed on a high continuum flux level in the X-ray as well as in the optical since a historic outburst in 2019. However, it shows
   strong short-term variations on timescales of a few months. 
    Densely sampled ASAS-SN V-band continuum data from 2014 till 2025 confirm that behavior. 
    \iras{} switched from a clear Seyfert 1 type in December 2019 to a  Seyfert 1.9/2 type in July 2020 based on its optical spectra. Afterward, it again became a Seyfert 1 type with symmetric broad single-peaked Balmer line profiles in January 2023.
    These spectra prove the repeating changing-look character of the galaxy.
  \iras{} exhibits extreme high Balmer decrements \Ha{}/\Hb{}  based on their broad line components. The Balmer decrement values are on the order of 10.
    \iras{} successively showed all types of broad line Balmer profiles during the past 25 years over periods of many years: asymmetric single-peaked, double-peaked, as well as single-peaked and symmetric profiles in addition to its Seyfert 1.9/2 transition. These variations are not clearly correlated with continuum and line intensity variations.} 
{}

   \maketitle
%

\section{Introduction}

Seyfert 1 galaxies and quasars are known to be variable in all frequency bands (from X-ray, UV, optical, to IR) on timescales of hours to decades. They show typical root mean square (rms) intensity variations of 10–20 percent in the optical spectral range on timescales of years \citep{rumbaugh18}. Some extreme active galactic nuclei (AGNs) even vary by a factor of 2 or more in their optical luminosities; for example, Mrk\,110 (\cite{bischoff99}, \cite{kollatschny01},  \cite{yin25} and references therein).
Some variable AGNs not only vary in their continuum and broad emission line intensities but also change their Seyfert type based on their varying (optical) spectral profiles.
 These optical changing-look (CL) AGNs exhibit transitions from type 1 to type 2 or vice versa on timescales of months to years 
 \citep{collin73, kollatschny85, lamassa15, ricci23, zeltyn24}.
 
More than 100 Seyfert galaxies and quasars are known to have changed their optical spectral type. Examples include 
NGC\,1566 \citep{pastroriza70, Oknyansky19, parker19, ochmann24},
NGC\,3516 (\citealt{collin73}, \citealt{mehdipour22}), NGC\,7603 \citep{tohline76,kollatschny00},  NGC\,4151 \citep{penston84}, Fairall\,9 \citep{kollatschny85}, NGC\,7582 \citep{aretxaga99}, NGC\,2617 \citep{shappee14}, Mrk\,590 \citep{denney14}, HE\,1136-2304  \citep{parker16, zetzl18, kollatschny18}, Mrk\,1018  \citep{mcelroy16,husemann16,kim18,lyu21},
NGC\,3516 \citep{popovic23}, and 1ES\,1927+654 (\citealt{trakhtenbrot19}, \citealt{ricci20},  \citealt{laha22}), and references therein. 
 Additional
 recent findings are based on spectral variations detected in large-scale surveys, such as the Sloan Digital Sky Survey \citep[SDSS; e.g.,][]{komossa08, lamassa15, rumbaugh18, macleod19, panda24, zeltyn24}, the Catalina Real-time Transient Survey \citep{graham20}, 
or the Wide-field Infrared Survey Explorer \citep{stern18}. 
A few of these CL AGNs have been found to change from Seyfert 1 type
to Seyfert 1.9/2 type and back to Seyfert 1 type or vice versa. This happens on timescales of a few years. \cite{wang24} and \cite{wang25} discuss investigations of approximately ten known recurrent CL AGNs.
For a recent review see \cite{komossa24}.

\iras{} was registered as a galaxy in the IRAS Faint Source Catalogue \citep{moshir90} in 1990.  Afterward it was classified as a Seyfert type 1 object in a spectroscopic survey in 1991 \citep{allen91}. An apparent magnitude of $13.68\pm 0.10$ in the r band \citep{shectman96} 
corresponds to an absolute magnitude of M$_{r} = -22.33 \pm 0.51$\,mag. 
\iras{} has been detected with ROSAT. Originally it was classified as a bright X-ray source \citep[ROSAT RASS catalog; e.g.,][]{grupe01}.
However, a very strong X-ray decline has been noticed within the \xmm{} slew survey in 2017. Subsequently, deep follow-up \swift{}, \xmm{}, and \textit{\nustar}{} observations in combination with optical spectra of \iras{} taken in 2017 showed  that \iras{} decreased in X-rays by a factor of more than 30
with respect to ROSAT and \swift{} data taken 10 to 27 years before \citep{kollatschny20}. 

A strong re-brightening by a factor of about 10 has been detected in the X-ray and UV continua of \iras{} with \swift{} in August 2019  \citep{grupe19}.
We then took follow-up X-ray observations together with optical spectroscopy of this AGN from 2019 until 2021 \citep{kollatschny23}.
\iras{} varied in the X-ray continuum by a factor of 4 and in the optical continuum by a factor of 1.6 within two months. The broad Balmer lines increased by similar amplitudes during the outburst in 2019. The \Ha{} and \Hb{} emission-lines showed broad double-peaked profiles during the minimum state in 2017. However, there were no major profile variations in the broad double-peaked Balmer profiles despite the strong intensity variations in 2019. One year after the outburst, \iras{} changed its optical spectral type and became a Seyfert 1.9/2  object in 2020.
Properties of \iras{} have been listed in 
 a recent overview of 20 changing-look AGNs with long-term optical and X-ray observations \citep{jana25}.

\begin{figure}
\centering
\vspace*{-32mm} 
\includegraphics[width=9.5cm,angle=0]{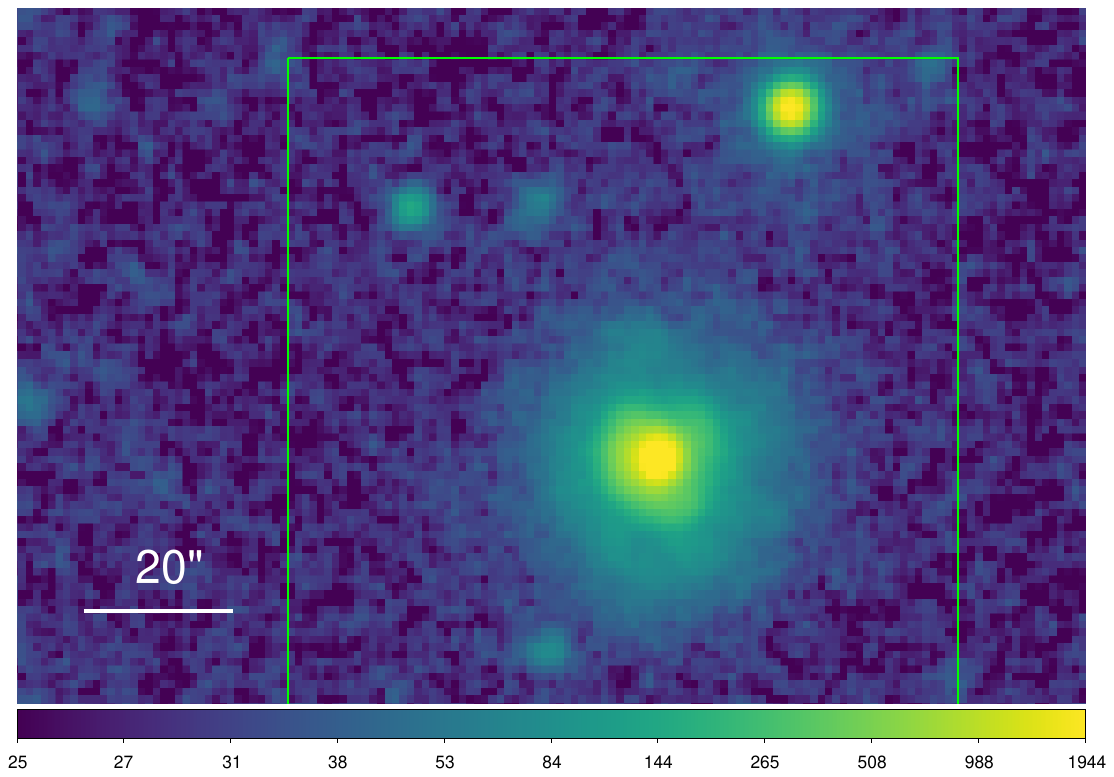}
\vspace*{-35mm} 
\caption{Integrated \swift{} image (25 ksec) in the UVOT-W2 band of \iras{}. North is to the top, east to the left. 20\,arcsec correspond to 13.7\,kpc at the distance of \iras{}.}
 \label{Im_SWIFT_iras_23226_w2}
\end{figure}

Figure~\ref{Im_SWIFT_iras_23226_w2} displays the \swift{} image  in the UVOT\,W2 band at 1928 \AA{} (see Sect.~\ref{sec:swift_observations}) that was merged from all W2 data. This image has a total
exposure time of 25 ks. 
The projected angular diameter of the galaxy is
about 40\,arcsec, which corresponds to a physical size of 
27\,kpc.

\begin{figure}
\centering
\includegraphics[width=6.cm,angle=0]{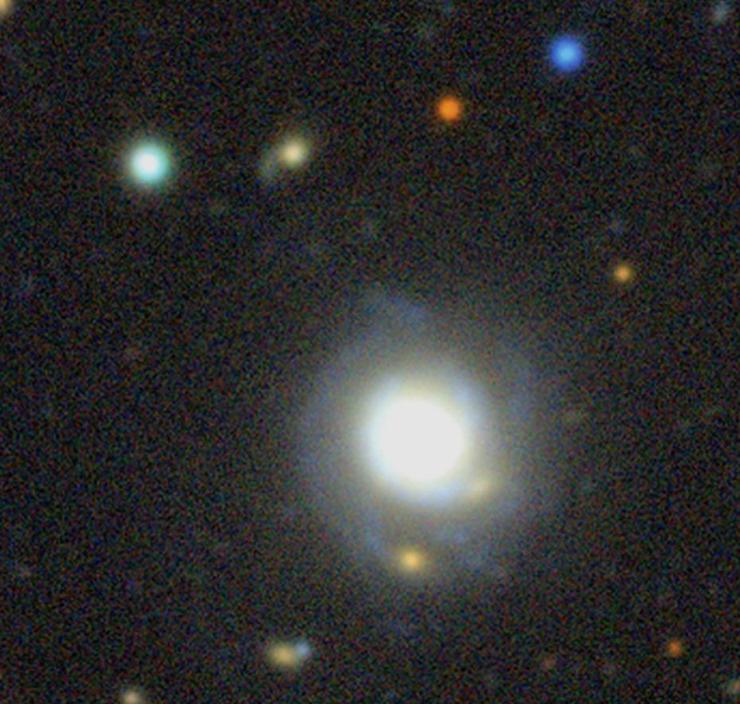}
\caption{DESI legacy image (g,r,i,z-band) of \iras{}. North is to the top, east to the left. This image shows the section from Fig.\,~\ref{Im_SWIFT_iras_23226_w2}.}
 \label{Im_Desi10_1_IRAS23226}
\end{figure}

The morphology of \iras{} has been
classified to be of the S0 type based on UKST sky survey images \citep{loveday96}. Figure~\ref{Im_Desi10_1_IRAS23226} displays a combined g,r,i,z-band DR10 image of \iras{} based on the 
Legacy Survey \citep{dey19}. This image shows indications of weak outer spiral arms.


Here we present new optical spectra taken in 2023 and 2024 showing that \iras{} switched back to a Seyfert 1 type. In addition we discuss new optical, UV, and 
X-ray variations over the past years based on \swift{} and ASAS-SN photometry. 

Throughout this paper, we assume a lambda cold dark matter ($\Lambda$CDM) cosmology with a Hubble constant of H$_0$~=~73~km s$^{-1}$ Mpc$^{-1}$,  $\Omega_{\Lambda}=0.72$, and $\Omega_{\rm M}=0.28$. With a redshift of {\bfseries $z=0.0359$}, this results in a luminosity distance of $D_{\rm L} = 144$\,Mpc for \iras{} ($\alpha_{2000} =$ 23h 25m 24.2s, $\delta = -38^{\circ}$ 26$^{'}$ 49.2$^{''}$) using the Cosmology Calculator developed by  \cite{wright06}.

\section{Observations and data reduction}

A dramatic increase in the X-ray and UV fluxes of \iras{} was discovered with \swift{} on August 11, 2019. We carried out follow-up continuum studies of \iras{} with \swift{} since that time.
 Furthermore, we took additional optical spectra with SALT as well as with the South African Astronomical Observatory (SAAO) 1.9\,m telescope to investigate the optical spectral variations \citep{kollatschny23}.

\subsection{\bf X-ray, UV, and optical continuum observations with \swift{}}
\label{sec:swift_observations}

\iras{} has been observed  with the NASA \textit{Neil Gehrels} \swift{} Gamma-Ray Burst Explorer Mission   \citep{gehrels04}  in the X-rays and 
UV/optical since 2007 at irregular intervals (see Fig.\,\ref{DirkLcIras23226_xrt_uvot_lc_20250520}
and Table\,\ref{swiftdata}).
After the detection of an increasing X-ray flux in August 2019, we started 
a detailed monitoring campaign of \iras{} with \swift{} in the X-rays and 
UV/optical from August 2019 until initially January 2020 \citep{kollatschny23}. 
After that we observed \iras{} with \swift{} at 19 epochs between April 2021 and August 2025.
The \swift{} observing dates and exposure times are listed in Table\,\ref{swiftdata}. 

All observations  with \swift's X-ray Telescope \citep[XRT;][]{burrows04} were performed in the photon counting mode \citep[pc mode;][]{hills05}. XRT data were processed using {\it xrtpipeline} 0.13.7, which is part of the HEARSOFT version 6.35.1 software package. 
Spectra were extracted in a circular region with a radius of 15 pixel (equal to 35$^{\prime\prime}$) for the source region and 100 pixel (equal to 235$^{\prime\prime}$) for the background region using the FTOOL {\it XSELECT}. Using {\it xrtmkarf}, an auxiliary response file (arf) was created and combined with the XRT response file {\it  swxpc0to12s6\_20210101v015.rmf}, except for the first two observations from 2007 that were performed before the change in the substrate voltage in the XRT detector in August 2007. For these two observations, the XRT response file {\it  swxpc0to12s0\_20070901v012.rmf} was applied. Note that these are new calibration files compared with \cite{kollatschny23}, and all data prior to that paper were reanalyzed. 
Due to the low number of counts in each observation, 
generally the spectra were not binned, and analyzed using w statistics \citep{cash79} within XSPEC version  12.15.0 \citep{arnaud96}. The count rates, the hardness ratios, the photon index $\Gamma$, the absorption-corrected 0.3-10\,keV flux, and the goodness of the fit are listed in Table\,\ref{swiftdata}. We define the hardness ratio as $HR = \frac{H-S}{H+S}$, where S are the counts in the 0.3-1.0 keV and H in the 1.0-10.0\,keV bands. Hardness ratios were determined applying  Bayesian statistics with the program {\it BEHR} by \citet{park06}.

UV and optical photometry was performed with \swift's 
UV-Optical Telescope \citep[UVOT;][]{roming05} in all six photometric filters\footnote{ UVW2 (1928 \AA), UVM2(2246  \AA), UVW1 (2600 \AA), U (3465 \AA), B (4392 \AA), and V (5468 \AA)}, except in some cases when the observations were interrupted by a \swift-detected gamma ray burst or by a high priority target-of-opportunity observation. 
Although most observations were performed in a single orbit, in some cases with a longer exposure time the observations had to be split over multiple orbits. All snapshots were merged using the tool {\it uvotimsum}. 
Source counts were extracted with a standard circular region with a radius of 5" and the background counts from a nearby circular region with an extraction radius of 20". 
The magnitudes listed in Table \ref{swiftmag} are in the Vega system.  These as well as the flux densities in each photometric filter were measured with the UVOT tool {\it uvotsource}  using the count rate conversion and calibration, in the manner described in \citet{poole08} and \citet{breeveld10}. 

The fluxes in each of the six UVOT filters are listed in Table~\ref{swiftmag} and were corrected for Galactic reddening. We determined the Galactic reddening correction for each filter by applying Eq. 2 in \citet{roming05}, who used the reddening curves published by \citet{cardelli89}. The attenuation in the direction of\,\iras{} is $E_{\rm B-V}=0.021$ \citep{schlafly11}. This results in the following magnitude corrections: UVW2$_{\rm corr} = 0.20$,  UVM2$_{\rm corr} = 0.244$,
UVW1$_{\rm corr} = 0.177$, U$_{\rm corr} = 0.136$, B$_{\rm corr} = 0.177$, and  V$_{\rm corr} = 0.083$. 

\subsection{\bf Continuum observations based on the All Sky Automated Survey for SuperNovae (ASAS-SN) photometry }
\label{sec:asas-sn}

We created another optical light curve of \iras{} using the All Sky Automated Survey for SuperNovae \citep[ASAS-SN\footnote{\url{https://www.astronomy.ohio-state.edu/asassn/}};][]{henden12} photometry from 2014 to 2025. 
ASAS-SN is a project designed to monitor the entire extragalactic sky with a mean cadence of $2-3$\,days to search  for transient events \citep[]{shappee14, kochanek17, jayasinghe18}. 
From the total light curve, we excluded epochs for which the source was not detected at a $5 \sigma$ confidence level. Subsequently, we removed outliers from the light curve by computing the rolling mean ($\Delta t = 30$\,days), discarding points outside of the $\pm 1 \sigma$ interval in each bin. Our aim was to ensure an appropriate balance between noise suppression and signal preservation for every interval of the light curve. Altogether, we removed $\sim$ 20\% of the original measurements. 

Because the ASAS-SN project employed both V-band and g-band filters, transitioning from V to g in 2017, there is a systematic offset between V-band and g-band observations of the source. In order to create a combined V-band + g-band light curve, we applied a flux shift to the g-band light curve to match the V-band light curve in the overlapping epochs (see, e.g., \citealt{kollatschny24}). To obtain the optimal shift parameter, we used the von Neumann algorithm \citep[]{neumann41} as a measure of noise for the combined light curve and chose the shift that minimized the noise in the overlapping epochs.

Our combined optical light curve of \iras{} is shown in Fig.~\ref{IRAS23226_ASAS-SN-14-25} lasting from April 30, 2014 (MJD 56777) until September 1, 2025 (MJD 60919). 
Altogether, we used photometric data for 4443 epochs.
The gaps in the light curve are caused by the annual observing break due to the object's position in the sky being too close to that of the Sun.

\subsection{Optical spectroscopy with the SAAO 1.9\,m telescope} \label{sec:SAAO_1.9}
\begin{table}
\centering
\tabcolsep+1.5mm
\caption{Log of spectroscopic observations of \iras{}.} 
\begin{tabular}{c c r l l}
\hline \hline
\noalign{\smallskip}
MJD & UT Date & \mcc{Exp. time} & Tel. & Obs. cond.\\
    &   &  \mcc{[s]} &  \\
    
\hline 
50724         &       1997-10-03   &       1800  & SAAO 1.9\,m & clear; 1\farcs{}8  \\

51350         &       1999-06-21      &      600  & Cerro-Tol. 4\,m   &   clouds; --  \\

57883        &       2017-05-10      &      900  & SALT  & clear; 1\farcs{}5  \\

57916         &       2017-06-12      &      900  & SALT & clear; 2\farcs{}0  \\ 

58727         &       2019-09-01      &    2400 & SAAO 1.9\,m   & clear; 1\farcs{}8\\

58736         &       2019-09-10      &     1200  & SALT & clouds; 2\farcs{}0 \\

58750         &       2019-09-24      &      900  & SALT & clear; 1\farcs{}5  \\

58794         &       2019-11-07      &      900  & SALT & clear; 1\farcs{}4  \\

58823         &       2019-12-06      &      900  & SALT & clear; 1\farcs{}3   \\

59053         &       2020-07-23      &    2400 & SAAO 1.9\,m   & clear;  1\farcs{}8 \\
59951         &       2023-01-06      &    2400  & SAAO 1.9\,m   & clear;  1\farcs{}8   \\

60635         &       2024-11-20      &    2400 & SAAO 1.9\,m   & clear;  1\farcs{}7 \\
\hline

\end{tabular}
\tablefoot{
The new observations have been taken with the SAAO 1.9\,m telescope in 2023 and 2024. In addition, we give the former observing dates already presented in  \cite{kollatschny23}.  We list the modified Julian date, the UT date, the exposure time, the telescope, and the observing conditions, including the seeing.}
\label{saltlog}
\end{table}

We took new optical spectra of \iras{} with the 1.9\,m telescope at the SAAO in Sutherland (South Africa) on January 6, 2023, and November 20, 2024.
We present the log of our spectroscopic observations in Table~\ref{saltlog}
together with former observing dates (1997 -- 2000) already presented in  \cite{kollatschny23}. Seeing was better than 2\arcsec and weather conditions were good.

We used the 
SpUpNIC spectrograph \citep{Crause19} with the low-resolution grating (300 lines/mm, dispersion
of 2.7 \AA/pixel) set at an angle to record the full optical range and with a slit width 
corresponding to 2\farcs{}7 on the sky. The spectra were extracted over 10\arcsec. The spectra were reduced in a homogeneous way with IRAF reduction packages.
The wavelength was determined by means of the emission spectrum of an argon lamp. The flux calibration was achieved through observations on the same night of spectrophotometric standard stars from the compilation of \citet[LTT\,9491, LTT\,7379, and LTT\,377]{Hamuy94}.  
The internal flux calibration was later adjusted so that the [\ion{O}{III}] emission line strengths corresponded to the same absolute [\ion{O}{iii}]\,$\lambda$5007 flux of $9.27 \times 10^{-15} \rm erg\,s^{-1}\,cm^{-2}$ as is presented in \cite{kollatschny23}.

\section{Results}

\subsection{\bf X-ray, UV, and optical continuum band variations based on \swift{}}
\label{sec:swift_results}
\begin{figure}
\centering
\vspace*{-11mm} 
\hspace*{-13mm} 
\includegraphics[width=11.cm,angle=0]{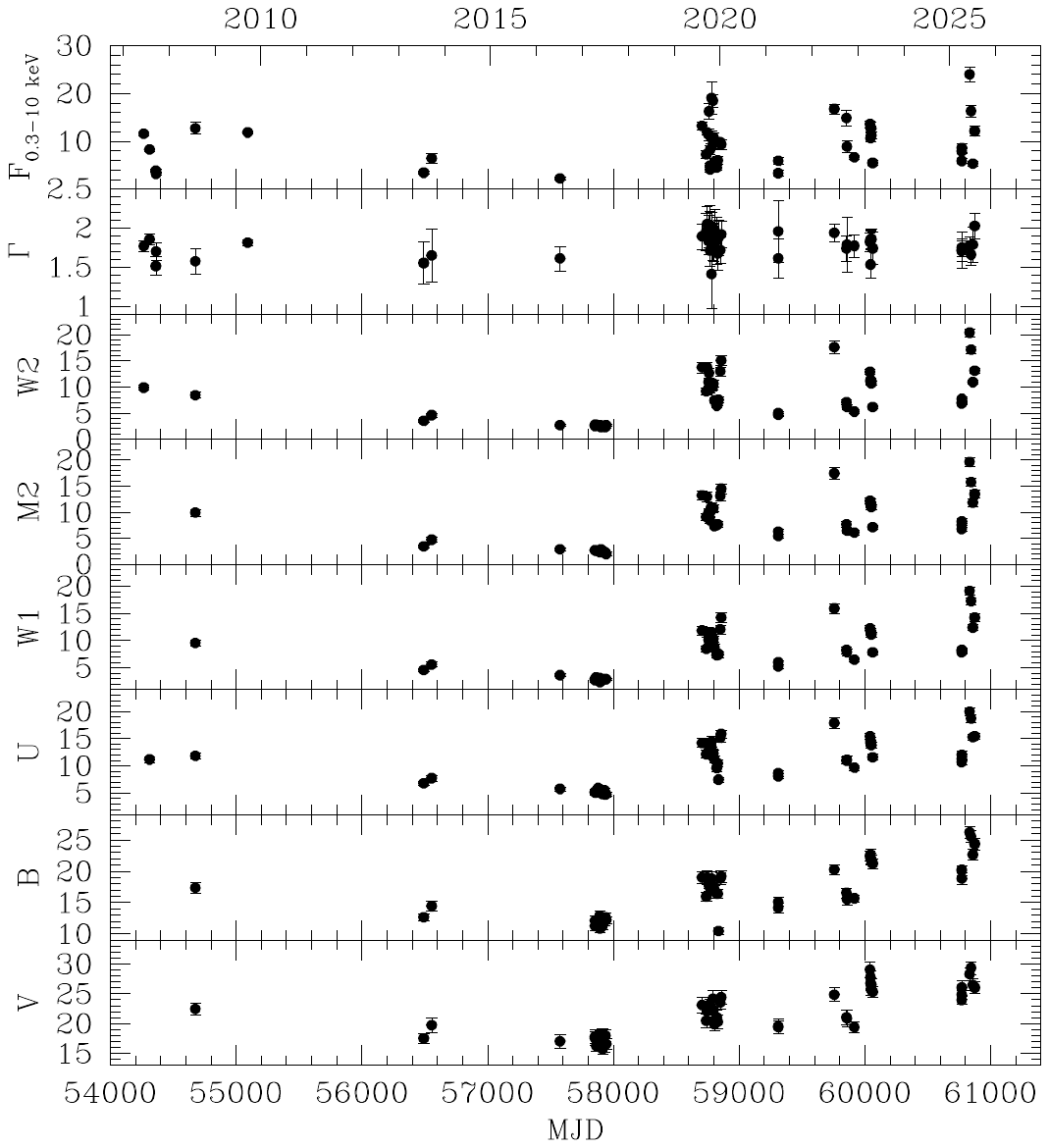} 
\vspace*{-33mm} 
\caption{Combined X-ray, UV, and optical light curves taken with the \swift{} satellite for the years 2007 until 2025. The fluxes are given in units of 10$^{-12}$ ergs s$^{-1}$ cm$^{-2}$.
$\Gamma$ is the X-ray photon index. }
\label{DirkLcIras23226_xrt_uvot_lc_20250520}
\end{figure}

We show in Fig.~\ref{DirkLcIras23226_xrt_uvot_lc_20250520}
the \swift{} 0.3--10\,keV flux and the X-ray photon index, as well as the UV and optical continuum light curves for the years 2007 to 2025. 
All of these measurements are listed in Tables~\ref{swiftdata} and
\ref{swiftmag}. 
A zoom-in of the light curves is presented in Fig.~\ref{DirkLcIras23226zoom_xrt_uvot_lc_20250819} for the years 2019 - 2025.
\begin{table}
\centering 
\caption{Variability statistics of the \swift{} continua  XRT, W2, M2, W1, U, B, and V (and their central wavelengths in \AA) for the years 2007 - 2025, 2007 -- 2017, and 2019 -- 2025: minimum flux, maximum flux, 
peak-to-peak ratio, mean flux, standard deviation, fractional variation, and their uncertainty.} 
\small
\begin{tabular}{lrrrrrrr} 
\hline 
\hline 
Cont. & $F_{\text{min}}$ & $F_{\text{max}}$ &  $R_{\text{max}}$ &  $\langle F \rangle$ & $\sigma_F$ & $F_{\text{var}}$ &  $\sigma_{F_{var}}$ \\ 
\hline 
'07--'25\\
\hline 
XRT  & $2.22$ & $23.96$ & $10.79$ & $9.71$ & $4.71$ & $0.47$ & $0.05$ \\
W2--1928& $2.29$ & $20.39$ & $8.90$ & $7.83$ & $4.59$ & $0.58$ & $0.06$ \\
M2--2246& $1.98$ & $19.63$ & $9.91$ & $7.82$ & $4.54$ & $0.58$ & $0.06$ \\
W1--2600& $2.31$ & $19.10$ & $8.27$ & $8.11$ & $4.23$ & $0.52$ & $0.05$ \\
U--3464& $4.69$ & $19.95$ & $4.25$ & $10.60$ & $4.13$ & $0.39$ & $0.04$ \\
B--4392& $4.68$ & $26.22$ & $5.60$ & $16.55$ & $4.49$ & $0.27$ & $0.03$ \\
V--5468& $15.7$ & $29.32$ & $1.87$ & $21.56$ & $3.83$ & $0.17$ & $0.02$ \\
\hline 
'07--'17\\ 
\hline 
XRT & $2.22$ & $12.71$ & $5.73$ & $7.04$ & $3.93$ & $0.55$ & $0.13$ \\
W2 & $2.29$ & $9.85$ & $4.30$ & $3.39$ & $2.05$ & $0.60$ & $0.10$ \\
M2 & $1.98$ & $9.92$ & $5.01$ & $3.09$ & $1.75$ & $0.56$ & $0.10$ \\
W1 & $2.31$ & $9.57$ & $4.14$ & $3.55$ & $1.64$ & $0.46$ & $0.08$ \\
U  & $4.69$ & $11.80$ & $2.52$ & $6.10$ & $1.98$ & $0.32$ & $0.05$ \\
B  & $4.68$ & $17.32$ & $3.70$ & $11.95$ & $2.26$ & $0.18$ & $0.03$ \\
V  & $15.7$ & $22.44$ & $1.43$ & $17.64$ & $1.45$ & $0.06$ & $0.02$ \\
\hline 
'19--'25\\ 
\hline 
XRT & $3.30$ & $23.96$ & $7.26$ & $10.34$ & $4.65$ & $0.44$ & $0.05$ \\
W2  & $4.62$ & $20.39$ & $4.41$ & $10.24$ & $3.72$ & $0.36$ & $0.04$ \\
M2  & $5.40$ & $19.63$ & $3.64$ & $10.40$ & $3.36$ & $0.32$ & $0.04$ \\
W1  & $5.26$ & $19.10$ & $3.63$ & $10.46$ & $3.07$ & $0.29$ & $0.04$ \\
U   & $7.44$ & $19.95$ & $2.68$ & $12.98$ & $2.76$ & $0.21$ & $0.03$ \\
B   & $10.4$ & $26.22$ & $2.51$ & $18.99$ & $3.32$ & $0.17$ & $0.02$ \\
V   & $19.4$ & $29.32$ & $1.51$ & $23.76$ & $2.88$ & $0.11$ & $0.02$ \\
\hline 
\hline 
\label{tab:contLightcurves07_25} 
\vspace{-.6cm}
\end{tabular} 
\flushleft 
\begin{tablenotes} 
\footnotesize 
\textbf{Notes:} In units of $10^{-12}\,$erg$\,$cm$^{-2}$s$^{-1}$. 
\end{tablenotes} 
\end{table}

Table~\ref{tab:contLightcurves07_25} gives the variability statistics based on the \swift{} continua (XRT, W2, M2, W1,
 U, B, and V). 
We give the minimum and maximum fluxes, $F_{\rm min}$ and $F_{\rm max}$, the peak-to-peak amplitudes, $R_{\rm max}$ = $F_{\rm max}$/$F_{\rm min}$,
the mean flux, $<$F$>\,=\frac{1}{N}\sum_{i=1}^{N}{F_i}$\,,\\
the standard deviation, $\sigma_{\rm F}$, with ${\sigma_{\rm F}}^2 = \frac{1}{N-1}\sum_{i=1}^{N}(F_i\,-<F>)^2$, and the fractional variation, $F_{\rm var}$=$\sqrt{{\sigma}^2-{\delta}^2}/{<F>}$, as defined by \cite{rodriguez97}.
$\delta$ is the mean square value of the uncertainties with each flux measurement:  $\delta^2 = \frac{1}{N}\sum_{i=1}^{N}{{\delta}^2_i}$.
The $F_{\rm var}$ uncertainties are defined in \cite{edelson02} as  $\sigma_{F_{var}}=\frac{1}{F_{var}}\sqrt{\frac{1}{2N}}\frac{{\sigma}^2}{{<F>}^2}$.
We present the variability statistics for the whole observing period from 2007 until 2025 (Table~\ref{tab:contLightcurves07_25}) as well as separately for the period before the outburst (2007--2017)  and after the outburst (2019--2025). 

We observe a clear decrease in the peak-to-peak amplitude and the fractional
variation with increasing wavelength.
The AGN has increased its X-ray flux by a factor of about 10 compared to the low-state observations. In the optical V band it was only a factor of 1.9.
The UV and optical \swift{} bands closely follow the X-ray light curve. 
There were no major changes in the X-ray photon index over the years 
despite the strong variations.
The mean fluxes in the individual bands and the standard deviations, $\sigma_F$, were larger by a factor of 1.5 after the outburst in 2019 in comparison to the values obtained before.

\subsection{\bf Optical continuum band variations based on ASAS-SN}
\begin{figure*}
\centering
\includegraphics[width=17.cm,angle=0]{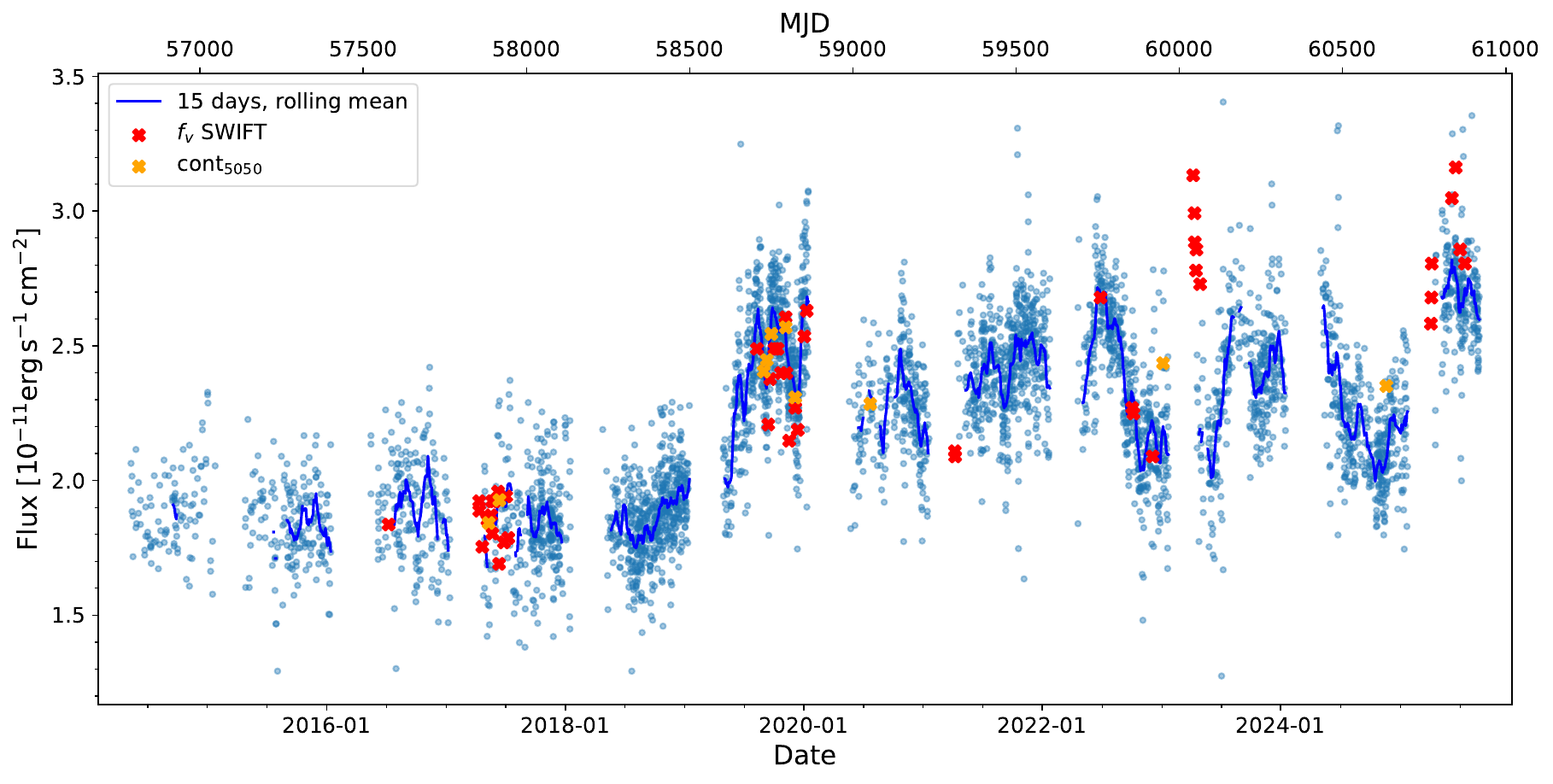}
\caption{Optical light curve of \iras{} based on ASAS-SN
data for the years 2014 to 2025. In addition we show the \swift{} V-band fluxes (red crosses) as well as the continuum fluxes at 5050\,\AA{} based on our optical spectra (orange crosses).
}
\label{IRAS23226_ASAS-SN-14-25}
\end{figure*}

Figure~\ref{IRAS23226_ASAS-SN-14-25} shows the optical light curve of \iras{} based on ASAS-SN data for the years 2014 to 2025. The ASAS-SN
fluxes have been inter-calibrated with respect to the V-band \swift{} fluxes.
In addition we show the inter-calibrated continuum fluxes at 5050\,\AA{} based
on our optical spectra.
The continuum flux remained relatively constant on a low level for the time period from 2014 up to January 2019. After that, the galaxy was not visible
for about four months because of the annual observing break due to the object's position in the sky being too close to that of the Sun.
After that gap -- until May 2019 -- we see that the continuum flux increased by a factor of about 1.5 until August 2019. After that the mean continuum flux remained more or less constant on this high level until August 2025. However, there were strong flux variations on timescales of months to years. 
We present the variability statistics of the V band ASAS-SN light curve
from 2014 until 2025, as well as for the subdivisions from 2014 to March 2019 before the outburst and March 2019 to 2025 after the outburst in 
Table~\ref{tab:cont19_25ASAS}.
\begin{table}[!htb] 
\centering 
\caption{Variability statistics of the V band ASAS-SN light curve
from 2014 until 2025, as well as for the subdivisions from 2014 to March 2019 before the outburst and March 2019 to 2025 after the outburst. } 
\begin{tabular}{lrrrrrr} 
\hline 
\hline 
ASAS V band  &  \multicolumn{1}{c}{$F_{\text{min}}$} &  \multicolumn{1}{c}{$F_{\text{max}}$} &  \multicolumn{1}{c}{$R_{\text{max}}$} &  \multicolumn{1}{c}{$\langle F \rangle$} &  \multicolumn{1}{c}{$\sigma_F$} &  \multicolumn{1}{c}{$F_{\text{var}}$} \\ 
\multicolumn{1}{c}{(1)} & \multicolumn{1}{c}{(2)} & \multicolumn{1}{c}{(3)} & \multicolumn{1}{c}{(4)} & \multicolumn{1}{c}{(5)} & \multicolumn{1}{c}{(6)} & \multicolumn{1}{c}{(7)} \\ 
\hline 
2014 -- 2025  & $1.27$ & $3.41$ & $2.67$ & $2.23$ & $0.33$ & $0.11$ \\
2014 -- 2019  & $1.29$ & $2.42$ & $1.87$ & $1.86$ & $0.16$ & $nan$ \\
2019 -- 2025  & $1.27$ & $3.41$ & $2.67$ & $2.39$ & $0.25$ & $0.07$ \\
\hline 
\hline 
\label{tab:cont19_25ASAS} 
\end{tabular} 
\footnotesize 
\tablefoot{Given are the minimum (2) and maximum flux density or integrated flux (3), peak-to-peak ratio (4), mean (5), standard deviation (6), and fractional variation (7). Continuum flux densities are in units of $10^{-11}\,$erg$\,$cm$^{-2}$s$^{-1}\,$\AA$^{-1}$.} 
\end{table}

\subsection{ Continuum and spectral line variations based on optical spectra}

The optical spectra of \iras{} were taken with different telescopes (see Table~\ref{saltlog}), different spectrographs, and different apertures. We recalibrated our spectra in a homogeneous way.
First we  inter-calibrated the spectral series taken with the two different apertures (SALT, SAAO).
The calibration  was carried out with respect to the [\ion{O}{iii}]\,$\lambda$4959,5007 lines, as well as with respect to the narrow \Ha{}, [\ion{N}{ii}], and [\ion{S}{ii}] line complex. The flux of the narrow emission lines is considered to be constant on timescales of years to decades. 
Afterward we inter-calibrated these two series with respect to the spectra that were taken nearly simultaneously in September 2019.
Finally, the spectra were calibrated to the same absolute [\ion{O}{iii}]\,$\lambda$5007 flux of $9.27 \times 10^{-15} \rm erg\,s^{-1}\,cm^{-2}$  taken on May 10, 2017, under clear conditions.

\begin{figure*}
\centering
\includegraphics[width=17cm,angle=0]{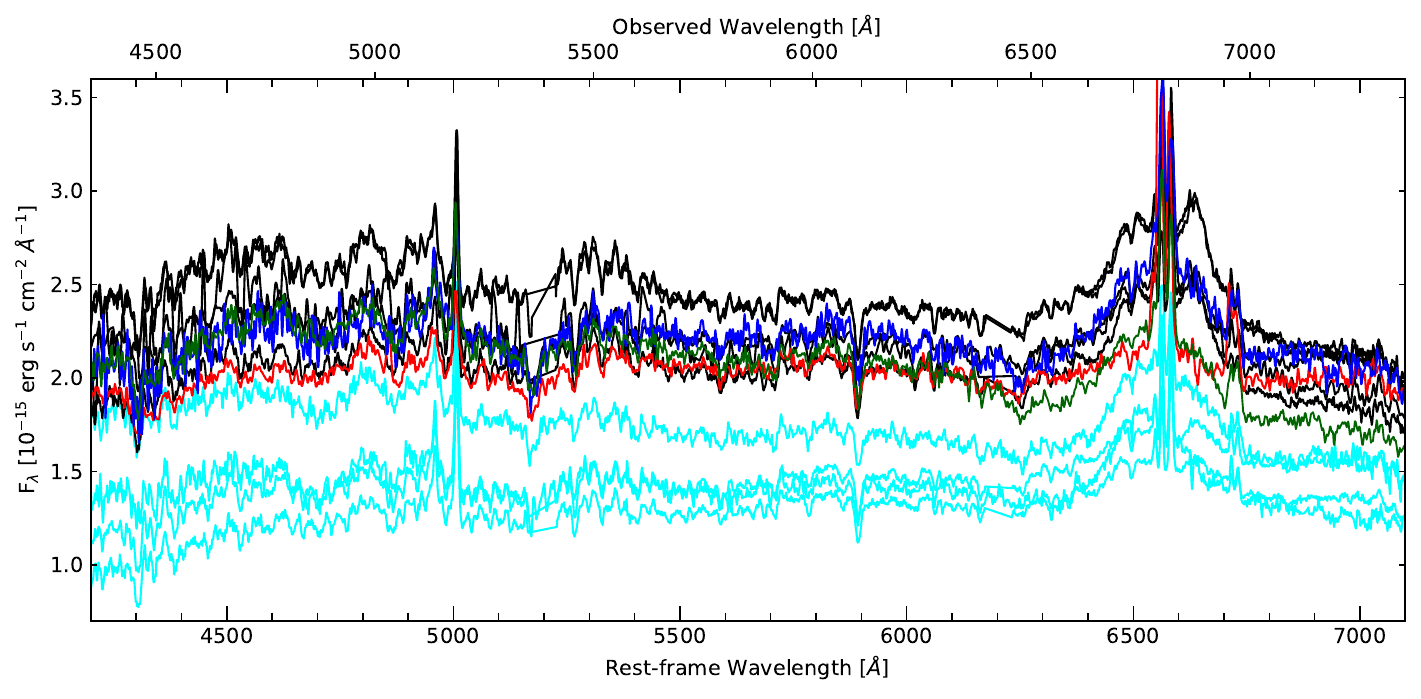}
\caption{Optical spectra of \iras{} for the years 1997 until 2024. The decreasing spectra from 1997 until 2017 are shown in cyan. The spectra taken during the outburst in 2019 are shown in black. The latest spectra are shown in red (2020), blue (2023), and green (2024).}
\label{IRAS23226_alleinteg_zuO3}
\end{figure*}

The optical spectra of \iras{}, taken at different epochs between 1997 and 2024, are presented in Fig.~\ref{IRAS23226_alleinteg_zuO3}.
The spectra that were taken from 1997 until 2017 and showing decreasing continuum 
flux are presented in cyan. The spectra taken during the outburst in 2019 are 
shown in black. The most recent spectra are presented in red (2020), blue (2023), and 
green (2024).

The obtained spectra contain a significant contribution from the underlying host
galaxy, as is indicated by the Mg and NaD absorption lines.
We carried out a spectral synthesis of the galaxy spectra using the penalized 
pixel-fitting method (pPXF; \citealt{cappellari04,cappellari17}) as is discussed in 
\cite{kollatschny23}. In this way we derived clean emission line fluxes and profiles after subtraction of the host galaxy contribution.

The broad emission line spectra are dominated by broad  \Hb{} and \Ha{} emission lines, as well as permitted broad \ion{Fe}{ii} blends of the multiplets 37, 38  and 42, 48, 49.  The wavelength ranges we used for the integration of the broad emission lines, as well as of the optical continuum  at 5050\AA{}, are given in Table~\ref{wavelength-ranges}. 
\begin{table}
\centering
\caption{Wavelength ranges for deriving the optical continuum, the Balmer lines, and \ion{Fe}{ii} blends in the rest frame.}
\label{wavelength-ranges}
\begin{tabular}{lcc}
\hline \hline
Continuum, Line & Wavelength range \\
           &    [\AA]         \\
\hline
Cont 5050    & 5040 -- 5060    \\
\hline
H$\beta_{\rm broad}$ & 4770 -- 5040   \\  
\ion{Fe}{ii}(42,48,49)   & 5230 -- 5520   \\
H$\alpha_{\rm broad}$ & 6240 -- 6900  \\  
\hline
\end{tabular}
\end{table}
The \ion{Fe}{ii} emission blend of the multiplets 42, 48, 49 begins at $\sim$5050\,\AA{}.
However, we start our flux measurements at 5230\AA{} because there is a gap in the SALT spectra and we want to derive the relative intensity variations.
We list in Table~\ref{BLR_variability} the continuum flux at $5050\pm20$\,\AA{}, as well as the broad-line fluxes of \Ha{} and \Hb{} and the \ion{Fe}{ii} blend (42, 48, and 49).  The broad H$\beta$ line has been corrected for the contribution of the narrow line fluxes of  H$\beta$ and  [\ion{O}{iii}]$\lambda\,4959,5007$. 
The broad H$\alpha$ line has been corrected for the contribution of the narrow line fluxes of H$\alpha$, [\ion{O}{i}]$\lambda$\,6300, [\ion{N}{ii}]$\lambda\lambda\,6548, 6584$, and [\ion{S}{ii}]$\lambda\lambda$\,6716, 6731. 
%
\begin{table*}
\centering
\caption{Fluxes of the optical continuum at $5050\pm10$\AA{}, 
the broad-line intensities of \Ha{} and \Hb{} after subtraction of the narrow line components, as well as 
the flux of the \ion{Fe}{ii} blend (42, 48, and 49) for $\lambda$ $\geq$ 5230\AA{}, and the BD \Ha{}/\Hb{}.}

\label{BLR_variability}
\begin{tabular}{lccccc}
\hline \hline
UT-Date & Cont$_{5050}$ & \Hb{} & \ion{Fe}{ii} & \Ha{} & \Ha{}/\Hb{} \\ 
\hline
1997-10-03   &   1.72    $\pm$   .28    &  27.2   $\pm$   4.  &  7.8    $\pm$   4.  &  161.5    $\pm$  17. &  5.9 $\pm$ 1.1\\
1999-06-21   &   1.38    $\pm$   .25    &  16.6   $\pm$   3.  &  13.6    $\pm$   4.  &  128.3   $\pm$  15. &  7.7 $\pm$ 1.7\\
2017-05-11   &   1.26    $\pm$   .19    &  11.3   $\pm$   3.  &  4.2    $\pm$   2.  &  121.8    $\pm$  14. & 10.8 $\pm$ 3.2\\
2017-06-13   &   1.40    $\pm$   .17    &  14.5   $\pm$   3.  &  4.6     $\pm$  2.  &  118.8    $\pm$  14. &  8.2 $\pm$ 2.0\\
2019-09-01   &   2.19    $\pm$   .35    &  15.9   $\pm$   4.  &  10.4    $\pm$   3.  &  134.0   $\pm$  19. &  8.4 $\pm$ 2.5\\
2019-09-10   &   2.26    $\pm$   .40    &  19.2   $\pm$   6.  &  21.1    $\pm$   5.  &  142.8   $\pm$  27. &  7.4 $\pm$ 2.8\\
2019-09-24   &   2.42    $\pm$   .18    &  18.7   $\pm$   3.  &  23.9    $\pm$   5.  &  213.6   $\pm$  12. & 11.4 $\pm$ 2.0\\
2019-11-07   &   2.46    $\pm$   .16    &  23.7   $\pm$   3.  &  31.0    $\pm$   5.  &  242.1   $\pm$  12. & 10.2 $\pm$ 1.4\\
2019-12-06   &   2.03    $\pm$   .14    &  12.7   $\pm$   3.  &  19.7    $\pm$   4.  &  231.4   $\pm$  12. & 18.2 $\pm$ 4.5\\
2020-07-23   &   1.99    $\pm$   .35    &   5.0  $\pm$    6.  &   9.9    $\pm$   6.  &   20.4   $\pm$  25. &  4.1 $\pm$ 7.0\\
2023-01-06   &   2.24    $\pm$   .24    &  11.7   $\pm$   4.  &  12.3    $\pm$  4.  &   165.3  $\pm$   17. & 14.1 $\pm$ 5.1\\
2024-11-20   &   2.10    $\pm$   .20    &  10.0   $\pm$   3.  &   9.9   $\pm$    4.  &  129.3  $\pm$   15. & 12.9 $\pm$ 4.2\\
\hline
\end{tabular}
\tablefoot{
Continuum flux densities are in units of 10$^{-15}$\,erg\,s$^{-1}$\,cm$^{-2}$\,\AA$^{-1}$. Line fluxes are in units of 10$^{-15}$\,erg\,s$^{-1}$\,cm$^{-2}$.
}
\end{table*} 
 \begin{figure*}
 \centering
  \includegraphics[width=1.0\linewidth,angle=0]{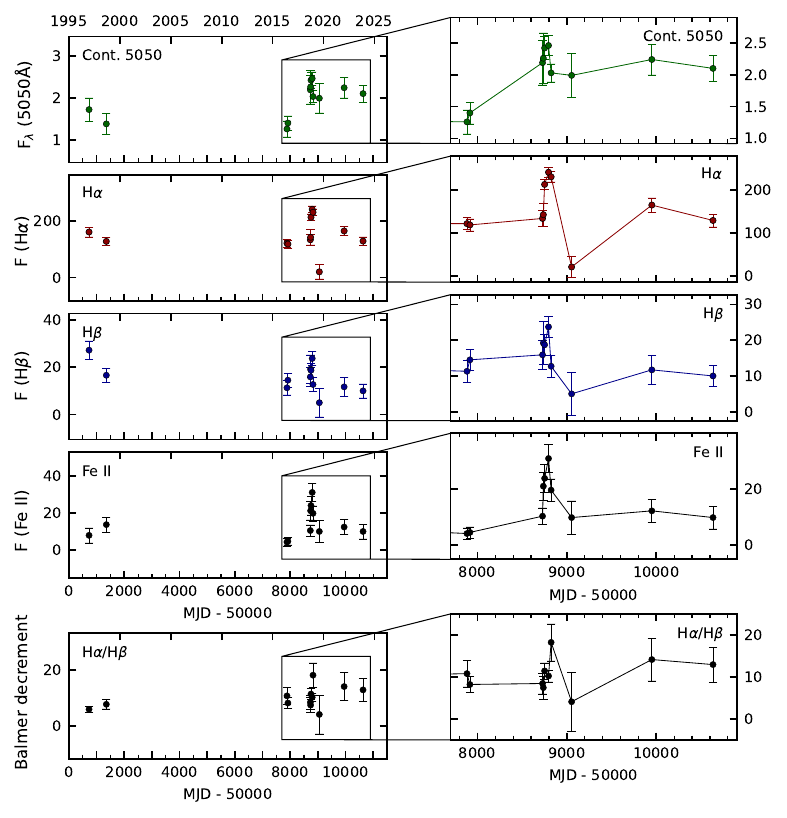}
     \caption{Long-term light curves of the continuum flux density at 5050\,\AA{} (in units of 10$^{-15}$\,erg\,s$^{-1}$\,cm$^{-2}$\,\AA$^{-1}$), as well as of the line fluxes of \Ha, \Hb, \ion{Fe}{ii}(42,48,49) (in units 10$^{-15}$\,erg\,s$^{-1}$\,cm$^{-2}$), and the BD for the years 1997 until 2024. The right panel shows the observations from 2017 to 2024 in more detail. 
     }
         \label{lc_IRAS23226_spectra_20250818.pdf}
   \end{figure*}
%
We present the long-term light curves of the
continuum flux at 5050\,\AA{}, as well as of the line fluxes of 
\Ha{}, \Hb{}, \ion{Fe}{ii}(42,48,49), and the Balmer decrement (BD) \Ha{}/\Hb{} for the epochs from 1997 until 2024 in Fig.~\ref{lc_IRAS23226_spectra_20250818.pdf}. The second panel shows the
variations 2017 to 2025 in more detail. 

\begin{figure}
\centering
\includegraphics[width=1.0\linewidth,angle=0]{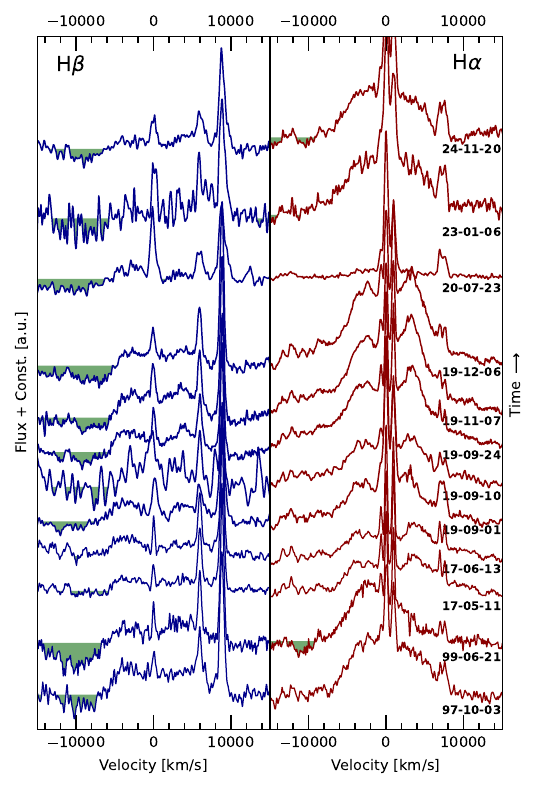}
\caption{Line profiles of \Ha{} and \Hb{} in velocity space
after subtraction of the host galaxy spectrum. Absorption components in the blue wing of the Balmer lines  (i.e., flux below zero) are shaded in green.}
\label{OchmIRAS23226_veloplots_20220325.pdf}
\end{figure}

The \Ha{} and \Hb{} emission-line profiles and their variations in velocity space are  presented in Fig.~\ref{OchmIRAS23226_veloplots_20220325.pdf} for the epochs 1997 until 2024. 
The early spectra taken in 1997 and 1999 show an asymmetrical profile where the blue wing is stronger than the red wing. Double-peaked profiles are present for the minimum continuum states from 2017 up to the maximum states in 2019.
The galaxy changed its type in 2020 when it became a Seyfert 1.9/2 type. The galaxy switched back to a Seyfert 1 type with symmetrical broad profiles for the years 2023 and 2024.

\section{Discussion}

\subsection{\bf Optical, UV, and X-ray continuum variations based on \swift{} and optical variations based on ASAS-SN}

The observed continuum variations taken with \swift{} show the same trend in all frequency bands (X-ray, UV, optical) from 2007 until 2025
(see Figs.~\ref{DirkLcIras23226_xrt_uvot_lc_20250520} and \ref{DirkLcIras23226zoom_xrt_uvot_lc_20250819}). There was a slow gradual decline of the X-ray and UV continuum fluxes by a factor of about five from 2007 until 2017. In 2019 an outburst occurred in all frequency bands: in the X-rays by a factor of about 5 and in the UV bands by a factor of about 4. 
Afterward we observe strong short-term variations by more than a factor of two on timescales of months. However, the average level -- over one to two years -- remained the same since the outburst in 2019  until the end of our campaign in September 2025. The variations in the optical U, B, and V bands were by a factor of 3 to 5 lower in comparison to the UV, X-ray variations. Here one must take into account that these bands were not corrected for the contribution of the host galaxy.

 We calculated the mean fluxes, $\langle F \rangle$, and their standard deviations, 
 $\sigma_F$, as well as the fractional variations, $F_{\rm var}$, for each \swift \ band in order to quantify the variability strengths.
The results are presented in Table~\ref{tab:contLightcurves07_25}. 
The fractional variations deduced from the \swift{} observations for the year 2007 until 2025 as well as for the periods 2007--2017 and 2019--2025 follow the same trend: the fractional variations decrease by a factor of about five from the X-ray to the optical bands. 
However, there were no major changes in the X-ray photon index -- despite the strong variations -- over all the years from 2007 until 2025.

We present in Fig.~\ref{IRAS23226_ASAS-SN-14-25} the denser sampled V-band variations based on ASAS-SN data for the years 2014 to 2025. 
They confirm the observed variations based on \swift{}. There were minor variations 
until the end of 2018. However, then the outburst happened in May 2019 with subsequent strong short-term variations on timescales of one to six months. The average flux value remained the same from 2019 until 2025  with $\langle F \rangle$ = $2.4 \times 10^{-11}$ ergs s$^{-1}$ cm$^{-2}$. 
The amplitudes of the short-term outbursts (30 -- 150 days) had typical values of
$\Delta F$ = $1 \times 10^{-11}$ ergs s$^{-1}$ cm$^{-2}$.
The variability statistics of the V-band ASAS-SN data is given in Table~\ref{tab:cont19_25ASAS}.
The V-band variability statistics (peak-to-peak ratio, standard deviation with regard to the mean, fractional variation) are similar for the \swift{} data and for the denser sampled ASAS-SN flux values.
It has already been mentioned in \cite{kollatschny23}
that the fractional variations in \iras{} are stronger by a factor of 2 in comparison to other highly variable changing-look AGNs such as HE\,1136-2304 \citep{zetzl18} or the variable AGN NGC\,5548 \citep{edelson15, fausnaugh16}.

\subsection{\bf The Eddington ratio}

Changing-look galaxies are known to often show low Eddington ratios, as Mrk\,1018 does for example \citep{lu25}. We want to test whether this also applies to \iras{}.
We calculated the Eddington ratio based on 
the Balmer lines obtained for the years 2023 and 2024 when the AGN was in a high state as well as based on the observed optical continuum intensities. The H$\alpha$ and H$\beta$ lines exhibit equal broad line widths  of 11 000 $\pm{}$ 300 \kms{} (full width at half maximum, FWHM).
This results in a  central black hole mass of $M=2.93\times 10^{8} M_{\odot}$  -- based on formulas given in \cite{trakhtenbrot12} -- and an Eddington luminosity
of $L_{Edd}$ = $4.39 \times 10^{46}$$\,$erg$\,$s$^{-1}$. Furthermore,
we measure a flux of $2.20\,\times 10^{-15}$\,erg\,s$^{-1}$\,cm$^{-2}$\,\AA$^{-1}$ at 5100 \AA{}, corresponding to 
 a luminosity, $L_{5100}$,  of $3.03 \times 10^{43}$$\,$erg$\,$s$^{-1}$.
With a bolometric correction (BC) according to \cite{netzer19} (where $k_{BOL}= c\times[L(\text{observed})/10^{42} \text{erg}\:\text{s}^{-1}]^d$ and c=\,40, d=\,-0.2), we derive 
a bolometric luminosity of $L_{bol}$ = $6.13 \times 10^{44}$$\,$erg$\,$s$^{-1}$.
This ultimately results in a low 
Eddington ratio of $L/L_\text{edd}$ = 0.014 for the years 2023 and 2024.
The continuum flux was 70 percent lower before the outburst in 2019. This gives an Eddington ratio of $L/L_\text{edd}$ = 0.010 for the time period from 1997 till 2017.

 This finding confirms that \iras{} belongs to the class of low Eddington AGNs. 
In addition, it is consistent with the median Eddington ratio, 0.01,  of a sample of 20 nearby optically identified changing-look AGNs \citep{jana25}. 
Furthermore, \cite{rumbaugh18} found a trend based on SDSS spectra that the Eddington ratio decreases in the case the maximum g-band variability increases.  
 
\subsection{\bf Continuum variability in \iras{} }

We observe a slow long-term continuum decrease in the optical, UV, and X-ray over many years from 1997 until 2019 based on the spectral and \swift{} continua, as well as on ASAS-SN data. 
The continuum flux  remained more or less constant on a low level for the  period 2014 to January 2019.  The galaxy was not visible for about four months after that period -- until May 2019 -- because of the annual observing break due to the object's position in the sky being too close to that of the Sun.
After that gap  we see an increase in the optical continuum flux by a factor of about 1.5 until August 2019. Since that outburst in 2019 the average flux remained more or less constant on its high level until the end of our campaign in August 2025. However, there were strong flux variations on timescales of months to years during that period. We see increasing or decreasing phases of the continuum flux on the order of 50 percent on short periods of 33 to 217 days (e.g., for the intervals MJD 58830 -- 58863 ( $\Delta$ = 33 d), MJD 59745 -- 59884 ( $\Delta$ = 139 d), MJD 60096 -- 60173 ( $\Delta$ = 77 d), and MJD 60626 -- 60843 ( $\Delta$ = 217 d)).
These repeated short-time variations are different from tidal disruption events (TDEs).
The majority of TDEs show single sharp flares \citep[e.g.,][]{vanvelzen21}. Only some of them show repeated high-amplitude flaring (tidal stripping events),  and essentially only the jetted events show additional rapid fluctuations \citep[e.g.,][]{komossa15, grupe24}. In any case, \iras{} is a classical AGN with a long-lived accretion disk and narrow-line region (NLR), with no need to invoke a TDE in the first place.

Furthermore, the strong short-term variations on timescales of months also exclude variable dust obscuration as their fundamental cause of variability. In addition, the X-ray hardness ratio remained more or less constant throughout the 18 years despite strong X-ray flux variations during that period. This is also an indication of no major dust absorption variability.  With a similar argument, \cite{lusso25} rule out that their observed variability behavior in the changing-look AGN NGC\,4614 is due to obscuration effects.
The different timescales to explain variability phenomena in AGNs have been listed and discussed in, for example, \citep{stern18}. 
This indicates that the general variability behavior in the continuum in \iras{} 
is very likely caused by varying accretion rates. It can explain the  short-term variations on timescales of months. The observed strong X-ray and optical continuum variations in \iras{} are also similar to those in the changing-look AGN HE\,1136-2304 \citep{zetzl18}.

\subsection{\bf Emission line intensity and profile variations}

The broad Balmer line intensity variations generally follow the continuum variations (see Fig.~\ref{lc_IRAS23226_spectra_20250818.pdf}).
 The Balmer line fluxes slightly decreased from the year 1997 until 2017. The fluxes increased by a factor of about two during the outburst in November 2019. 
Afterward they decreased again to a flux level before the outburst in 2019
and remained more or less constant until the end of 2024.
However, the final decreasing phase went into a lower state in the Balmer lines in comparison to the continuum.

In addition to the intensity variations, 
\iras{}  also showed line profile 
variations independent of the line and continuum fluxes over our observing period from 1997 to 2024 (see 
Fig.~\ref{OchmIRAS23226_veloplots_20220325.pdf}). 
These line profile variations are more clearly visible in H$\alpha$ in comparison to  H$\beta$. This is 
because of the far higher signal-to-noise ratios in the H$\alpha$ lines and the clearer host-galaxy subtraction for that line.
We observe three different line profile shapes, namely a blue asymmetry at the beginning, afterward double-peaked profiles independent of the continuum and line intensity fluxes, and finally symmetric emission line profiles.
\iras{} was an obvious Seyfert 1 type for the years 1997 and 1999. 
However, the profiles were asymmetric, with a stronger blue peak 
originating at a velocity of $-3100\pm500$\,\kms{}. 
It showed almost symmetrical double-peaked profiles with maxima at 
$\pm3100\pm300$\,\kms{} when it was 
observed in a low state in 2017. Additional spectra taken in 2019 
again show similar double-peaked profiles with maxima at 
$\pm3100\pm300$\,\kms{}. However the line fluxes were significantly 
stronger in 2019 in comparison to 2017 (see 
Fig.~\ref{lc_IRAS23226_spectra_20250818.pdf}).
The Balmer line intensities increased by factors of two at that time. Afterward, the spectral type 
suddenly changed to that of a Seyfert 1.9/2 type in 2020. This 
transition to a Seyfert 1.9/2 type happened without an observed major 
continuum decline.
Subsequently, \iras{} again became a clear Seyfert 1 
with single-peaked symmetrical line profiles over the period from January 2023 until November 2024.

\iras{} always showed very broad H$\alpha$ and H$\beta$ emission 
lines with quite similar line widths (FWHM) of 
11\,000$\pm{}$300\kms{} independent of the profile shape -- except for the short intermezzo when it became a Seyfert 1.9/2 type.
The existence of very broad line widths in this highly variable AGN is consistent with  the general finding that there is a correlation between emission line widths 
and the strength of their optical continuum variability amplitudes in AGNs \citep{kollatschny06}.

\iras{} exhibits extreme high BDs \Ha{}/\Hb{} based on their broad line components. The BD values are on the order of 10 (between 5.9 and 18.2; see Table~\ref{BLR_variability}). 
We show the variation of the BD in Fig.~\ref{lc_IRAS23226_spectra_20250818.pdf}.
It closely follows the H$\alpha$ flux variations.
Relatively high BD values are related
to lower continuum intensity phases in variable AGNs
(\citealt{kollatschny00, kollatschny18, kollatschny22}).
Typical BD values are between 3.2 and 4.2 in variable AGNs such as Mrk\,110 \citep{bischoff99} and Mrk\,926 \citep{kollatschny22}.
However, the BDs reach high values of up to 6.9 in low-intensity phases of highly variable AGNs such as NGC\,7603 \citep{kollatschny00} or even 7.4 in the changing-look AGN HE1136-2304.
These high BD values are probably caused by optical depth effects
\citep{korista04} in the inner emission line clouds.

\iras{} is a repeating changing-look AGN based on its emission line profile variations. It varied from a clear Seyfert 1 type to a Seyfert 1.9/2 type and back to a clear Seyfert 1 type. Only a few other AGNs such as Mrk\,1018
\citep{lu25} are known to be repeating changing-look AGNs
\citep{jana25, lusso25, ward25}.
The line profiles in \iras{} varied independently of the
continuum and line intensity fluxes and their variations. 
Such a behavior has been observed before in
the highly variable changing-look AGN HE\,1136-2304 \citep{kollatschny18}.

The variations in the ionizing continuum flux can explain the variations in the broad line intensities.
However, we see independent variations in the line profiles.
These line profile variations might be caused by variations in the accretion disk geometry or the accretion disk orientation (e.g.,  \cite{storchi97, kaaz23}, and references therein). In addition the emission line cloud distribution or their shapes might have changed on timescales of years.  
Furthermore, variable dust obscuration might have been taken on different timescales. 
 The profiles varied on timescales of several years, while the line intensities varied over shorter timescales for the years 2017 until 2024. We do not observe a straightforward 
 correlation between an increasing or decreasing continuum flux and line profile variations.

\section{Summary}

We present results on long- and short-term variations of the continuum in the X-rays, UV, and optical as well as spectral line-profile variations of the repeating changing-look AGN \iras{}.  Our findings are summarized below.

\begin{enumerate}[(1)]
 \item  
\iras{} showed strong X-ray, UV, and optical continuum variations
based on \swift{} observations taken between 2007 and 2025.
The light curves in the different bands show a similar variability pattern: 
 There was a slow gradual decline of the X-ray and UV continuum fluxes by about a factor of about five from 2007 until 2017. In 2019 an outburst occurred in all frequency bands: in the X-rays by a factor of about 6 and in the UV bands by a factor of about 4. 
Afterward we observe strong short term variations by more than a factor of two on timescales of months. The variations in the optical U, B, and V bands were lower by a factor of 3 to 5.

\item However, the average flux level -- over periods of one to five years -- remained constant in the X-rays, as  well as UV and optical bands, from the outburst in 2019  until the preliminary end of our campaign in September 2025. Denser sampled  ASAS-SN  V-band continuum observations from 2014 till 2025 confirm the \swift{} results.
  
\item The Balmer emission line profiles of \iras{} varied from single blue-peaked profiles in the years 1997 and 1999 to  double-peaked profiles for the years 2017 to 2019. The profiles of the double-peaked emission lines remained constant independently of line flux and continuum variations. \iras{} suddenly changed from a Seyfert type 1 to a Seyfert 1.9/2 object in 2022. Afterward it again became a Seyfert 1 type with symmetric broad emission profiles. This demonstrates that \iras{} is a repeating changing-look AGN.

\item \iras{} exhibits extreme high BDs \Ha{}/\Hb{}  based on their broad line components. These BD values are on the order of 10.

\item The continuum and their related  emission line intensity variations  in \iras{} are most probably caused by changes in the accretion rate. That is based on the observed short-term variations on timescales of weeks to months. 
 Furthermore, we observe  line-profile variations on timescales of several years that are not unambiguously correlated with increasing or decreasing  continuum flux variations. These changes might be explained by
changes in the accretion disk structure, their orientation, variable absorption, and/or the emission line cloud shapes or their distribution. 

\end{enumerate}

\iras{} is an interesting repeating changing-look AGN due to its strong variations in the optical and X-ray continuum, as well as its clear transitions from a single-peaked to a double-peaked Seyfert 1 type object, to a Seyfert 1.9/2 type, and finally to a symmetric Seyfert 1 type.

\begin{acknowledgements}
This work has been supported by the DFG grant KO 857/35-2. MWO acknowledges the support of the German Aerospace Center (DLR) within the framework of the ``Verbundforschung Astronomie und Astrophysik'' through grant 50OR2305 with funds from the BMWK. SK would like to thank the CAS President's International Fellowship Initiative for visiting scientists.
The paper includes observations obtained at SAAO.
We would like to thank  \swift\ PI Brad Cenko for approving our continued requests to observe IRAS\,23226-3843 and the \swift\ Science Operations team for executing the observations. 
This research has made use of the NASA/IPAC 
Extragalactic Database (NED) which is operated by the Jet Propulsion Laboratory, Caltech, under contract with the National Aeronautics and Space Administration. 
This work made use of data supplied by the UK Swift Science Data Centre at the University of Leicester \citep{evans07}. 
This research has made use of the XRT Data Analysis Software (XRTDAS) developed under the responsibility of the ASI Science Data Center (ASDC), Italy. This research has made use of data obtained through the High Energy 
Astrophysics Science Archive Research Center Online Service, provided by the NASA/Goddard Space Flight Center. 
This paper includes a Legacy Survey image. The DESI Legacy Imaging Surveys consist of three individual and complementary projects: the Dark Energy Camera Legacy Survey (DECaLS), the Beijing-Arizona Sky Survey (BASS), and the Mayall z-band Legacy Survey (MzLS). DECaLS, BASS and MzLS together include data obtained, respectively, at the Blanco telescope, Cerro Tololo Inter-American Observatory, NSF’s NOIRLab; the Bok telescope, Steward Observatory, University of Arizona; and the Mayall telescope, Kitt Peak National Observatory, NOIRLab. 

\end{acknowledgements}

\bibliographystyle{aa} 
\bibliography{literature} 

\begin{thebibliography}{88}
\expandafter\ifx\csname natexlab\endcsname\relax\def\natexlab#1{#1}\fi

\bibitem[{{Allen} {et~al.}(1991){Allen}, {Norris}, {Meadows}, \&
  {Roche}}]{allen91}
{Allen}, D.~A., {Norris}, R.~P., {Meadows}, V.~S., \& {Roche}, P.~F. 1991,
  \mnras, 248, 528

\bibitem[{{Aretxaga} {et~al.}(1999){Aretxaga}, {Joguet}, {Kunth}, {Melnick}, \&
  {Terlevich}}]{aretxaga99}
{Aretxaga}, I., {Joguet}, B., {Kunth}, D., {Melnick}, J., \& {Terlevich}, R.~J.
  1999, \apjl, 519, L123

\bibitem[{{Arnaud}(1996)}]{arnaud96}
{Arnaud}, K.~A. 1996, in Astronomical Society of the Pacific Conference Series,
  Vol. 101, Astronomical Data Analysis Software and Systems V, ed. G.~H.
  {Jacoby} \& J.~{Barnes}, 17

\bibitem[{{Bischoff} \& {Kollatschny}(1999)}]{bischoff99}
{Bischoff}, K. \& {Kollatschny}, W. 1999, \aap, 345, 49

\bibitem[{{Breeveld} {et~al.}(2010){Breeveld}, {Curran}, {Hoversten}, {Koch},
  {Landsman}, {Marshall}, {Page}, {Poole}, {Roming}, {Smith}, {Still},
  {Yershov}, {Blustin}, {Brown}, {Gronwall}, {Holland}, {Kuin}, {McGowan},
  {Rosen}, {Boyd}, {Broos}, {Carter}, {Chester}, {Hancock}, {Huckle}, {Immler},
  {Ivanushkina}, {Kennedy}, {Mason}, {Morgan}, {Oates}, {de Pasquale},
  {Schady}, {Siegel}, \& {vanden Berk}}]{breeveld10}
{Breeveld}, A.~A., {Curran}, P.~A., {Hoversten}, E.~A., {et~al.} 2010, \mnras,
  406, 1687

\bibitem[{{Burrows} {et~al.}(2005){Burrows}, {Hill}, {Nousek}, {Kennea},
  {Wells}, {Osborne}, {Abbey}, {Beardmore}, {Mukerjee}, {Short}, {Chincarini},
  {Campana}, {Citterio}, {Moretti}, {Pagani}, {Tagliaferri}, {Giommi},
  {Capalbi}, {Tamburelli}, {Angelini}, {Cusumano}, {Br{\"a}uninger}, {Burkert},
  \& {Hartner}}]{burrows04}
{Burrows}, D.~N., {Hill}, J.~E., {Nousek}, J.~A., {et~al.} 2005, \ssr, 120, 165

\bibitem[{{Cappellari}(2017)}]{cappellari17}
{Cappellari}, M. 2017, \mnras, 466, 798

\bibitem[{{Cappellari} \& {Emsellem}(2004)}]{cappellari04}
{Cappellari}, M. \& {Emsellem}, E. 2004, \pasp, 116, 138

\bibitem[{{Cardelli} {et~al.}(1989){Cardelli}, {Clayton}, \&
  {Mathis}}]{cardelli89}
{Cardelli}, J.~A., {Clayton}, G.~C., \& {Mathis}, J.~S. 1989, \apj, 345, 245

\bibitem[{{Cash}(1979)}]{cash79}
{Cash}, W. 1979, \apj, 228, 939

\bibitem[{{Collin-Souffrin} {et~al.}(1973){Collin-Souffrin}, {Alloin}, \&
  {Andrillat}}]{collin73}
{Collin-Souffrin}, S., {Alloin}, D., \& {Andrillat}, Y. 1973, \aap, 22, 343

\bibitem[{{Crause} {et~al.}(2019){Crause}, {Gilbank}, {Gend}, {Worters},
  {Sass}, {Kotze}, {Potter}, {Sickafoose}, {Sefako}, {Southworth}, {Macri},
  {Thorstensen}, {Galan}, {Skelton}, {Engelbrecht}, {Braker}, {Winkler},
  {Pie{\'n}kowski}, {S{\"u}rgit}, {Erdem}, \& {Burleigh}}]{Crause19}
{Crause}, L.~A., {Gilbank}, D., {Gend}, C.~v., {et~al.} 2019, Journal of
  Astronomical Telescopes, Instruments, and Systems, 5, 024007

\bibitem[{{Denney} {et~al.}(2014){Denney}, {De Rosa}, {Croxall}, {Gupta},
  {Bentz}, {Fausnaugh}, {Grier}, {Martini}, {Mathur}, {Peterson}, {Pogge}, \&
  {Shappee}}]{denney14}
{Denney}, K.~D., {De Rosa}, G., {Croxall}, K., {et~al.} 2014, \apj, 796, 134

\bibitem[{{Dey} {et~al.}(2019){Dey}, {Schlegel}, {Lang}, {Blum}, {Burleigh},
  {Fan}, {Findlay}, {Finkbeiner}, {Herrera}, {Juneau}, {Landriau}, {Levi},
  {McGreer}, {Meisner}, {Myers}, {Moustakas}, {Nugent}, {Patej}, {Schlafly},
  {Walker}, {Valdes}, {Weaver}, {Y{\`e}che}, {Zou}, {Zhou}, {Abareshi},
  {Abbott}, {Abolfathi}, {Aguilera}, {Alam}, {Allen}, {Alvarez}, {Annis},
  {Ansarinejad}, {Aubert}, {Beechert}, {Bell}, {BenZvi}, {Beutler}, {Bielby},
  {Bolton}, {Brice{\~n}o}, {Buckley-Geer}, {Butler}, {Calamida}, {Carlberg},
  {Carter}, {Casas}, {Castander}, {Choi}, {Comparat}, {Cukanovaite}, {Delubac},
  {DeVries}, {Dey}, {Dhungana}, {Dickinson}, {Ding}, {Donaldson}, {Duan},
  {Duckworth}, {Eftekharzadeh}, {Eisenstein}, {Etourneau}, {Fagrelius},
  {Farihi}, {Fitzpatrick}, {Font-Ribera}, {Fulmer}, {G{\"a}nsicke},
  {Gaztanaga}, {George}, {Gerdes}, {Gontcho}, {Gorgoni}, {Green}, {Guy},
  {Harmer}, {Hernandez}, {Honscheid}, {Huang}, {James}, {Jannuzi}, {Jiang},
  {Joyce}, {Karcher}, {Karkar}, {Kehoe}, {Kneib}, {Kueter-Young}, {Lan},
  {Lauer}, {Le Guillou}, {Le Van Suu}, {Lee}, {Lesser}, {Perreault Levasseur},
  {Li}, {Mann}, {Marshall}, {Mart{\'\i}nez-V{\'a}zquez}, {Martini}, {du Mas des
  Bourboux}, {McManus}, {Meier}, {M{\'e}nard}, {Metcalfe},
  {Mu{\~n}oz-Guti{\'e}rrez}, {Najita}, {Napier}, {Narayan}, {Newman}, {Nie},
  {Nord}, {Norman}, {Olsen}, {Paat}, {Palanque-Delabrouille}, {Peng},
  {Poppett}, {Poremba}, {Prakash}, {Rabinowitz}, {Raichoor}, {Rezaie},
  {Robertson}, {Roe}, {Ross}, {Ross}, {Rudnick}, {Safonova}, {Saha},
  {S{\'a}nchez}, {Savary}, {Schweiker}, {Scott}, {Seo}, {Shan}, {Silva},
  {Slepian}, {Soto}, {Sprayberry}, {Staten}, {Stillman}, {Stupak}, {Summers},
  {Sien Tie}, {Tirado}, {Vargas-Maga{\~n}a}, {Vivas}, {Wechsler}, {Williams},
  {Yang}, {Yang}, {Yapici}, {Zaritsky}, {Zenteno}, {Zhang}, {Zhang}, {Zhou}, \&
  {Zhou}}]{dey19}
{Dey}, A., {Schlegel}, D.~J., {Lang}, D., {et~al.} 2019, \aj, 157, 168

\bibitem[{{Edelson} {et~al.}(2015){Edelson}, {Gelbord}, {Horne}, {McHardy},
  {Peterson}, {Ar{\'e}valo}, {Breeveld}, {De Rosa}, {Evans}, {Goad}, {Kriss},
  {Brandt}, {Gehrels}, {Grupe}, {Kennea}, {Kochanek}, {Nousek}, {Papadakis},
  {Siegel}, {Starkey}, {Uttley}, {Vaughan}, {Young}, {Barth}, {Bentz},
  {Brewer}, {Crenshaw}, {Dalla Bont{\`a}}, {De Lorenzo-C{\'a}ceres}, {Denney},
  {Dietrich}, {Ely}, {Fausnaugh}, {Grier}, {Hall}, {Kaastra}, {Kelly},
  {Korista}, {Lira}, {Mathur}, {Netzer}, {Pancoast}, {Pei}, {Pogge},
  {Schimoia}, {Treu}, {Vestergaard}, {Villforth}, {Yan}, \& {Zu}}]{edelson15}
{Edelson}, R., {Gelbord}, J.~M., {Horne}, K., {et~al.} 2015, \apj, 806, 129

\bibitem[{{Edelson} {et~al.}(2002){Edelson}, {Turner}, {Pounds}, {Vaughan},
  {Markowitz}, {Marshall}, {Dobbie}, \& {Warwick}}]{edelson02}
{Edelson}, R., {Turner}, T.~J., {Pounds}, K., {et~al.} 2002, \apj, 568, 610

\bibitem[{{Evans} {et~al.}(2007){Evans}, {Beardmore}, {Page}, {Tyler},
  {Osborne}, {Goad}, {O'Brien}, {Vetere}, {Racusin}, {Morris}, {Burrows},
  {Capalbi}, {Perri}, {Gehrels}, \& {Romano}}]{evans07}
{Evans}, P.~A., {Beardmore}, A.~P., {Page}, K.~L., {et~al.} 2007, \aap, 469,
  379

\bibitem[{{Fausnaugh} {et~al.}(2016){Fausnaugh}, {Denney}, {Barth}, {Bentz},
  {Bottorff}, {Carini}, {Croxall}, {De Rosa}, {Goad}, {Horne}, {Joner},
  {Kaspi}, {Kim}, {Klimanov}, {Kochanek}, {Leonard}, {Netzer}, {Peterson},
  {Schn{\"u}lle}, {Sergeev}, {Vestergaard}, {Zheng}, {Zu}, {Anderson},
  {Ar{\'e}valo}, {Bazhaw}, {Borman}, {Boroson}, {Brandt}, {Breeveld}, {Brewer},
  {Cackett}, {Crenshaw}, {Dalla Bont{\`a}}, {De Lorenzo-C{\'a}ceres},
  {Dietrich}, {Edelson}, {Efimova}, {Ely}, {Evans}, {Filippenko}, {Flatland},
  {Gehrels}, {Geier}, {Gelbord}, {Gonzalez}, {Gorjian}, {Grier}, {Grupe},
  {Hall}, {Hicks}, {Horenstein}, {Hutchison}, {Im}, {Jensen}, {Jones},
  {Kaastra}, {Kelly}, {Kennea}, {Kim}, {Korista}, {Kriss}, {Lee}, {Lira},
  {MacInnis}, {Manne-Nicholas}, {Mathur}, {McHardy}, {Montouri}, {Musso},
  {Nazarov}, {Norris}, {Nousek}, {Okhmat}, {Pancoast}, {Papadakis}, {Parks},
  {Pei}, {Pogge}, {Pott}, {Rafter}, {Rix}, {Saylor}, {Schimoia}, {Siegel},
  {Spencer}, {Starkey}, {Sung}, {Teems}, {Treu}, {Turner}, {Uttley},
  {Villforth}, {Weiss}, {Woo}, {Yan}, \& {Young}}]{fausnaugh16}
{Fausnaugh}, M.~M., {Denney}, K.~D., {Barth}, A.~J., {et~al.} 2016, \apj, 821,
  56

\bibitem[{{Gehrels} {et~al.}(2004){Gehrels}, {Chincarini}, {Giommi}, {Mason},
  {Nousek}, {Wells}, {White}, {Barthelmy}, {Burrows}, {Cominsky}, {Hurley},
  {Marshall}, {M{\'e}sz{\'a}ros}, {Roming}, {Angelini}, {Barbier}, {Belloni},
  {Campana}, {Caraveo}, {Chester}, {Citterio}, {Cline}, {Cropper}, {Cummings},
  {Dean}, {Feigelson}, {Fenimore}, {Frail}, {Fruchter}, {Garmire}, {Gendreau},
  {Ghisellini}, {Greiner}, {Hill}, {Hunsberger}, {Krimm}, {Kulkarni}, {Kumar},
  {Lebrun}, {Lloyd-Ronning}, {Markwardt}, {Mattson}, {Mushotzky}, {Norris},
  {Osborne}, {Paczynski}, {Palmer}, {Park}, {Parsons}, {Paul}, {Rees},
  {Reynolds}, {Rhoads}, {Sasseen}, {Schaefer}, {Short}, {Smale}, {Smith},
  {Stella}, {Tagliaferri}, {Takahashi}, {Tashiro}, {Townsley}, {Tueller},
  {Turner}, {Vietri}, {Voges}, {Ward}, {Willingale}, {Zerbi}, \&
  {Zhang}}]{gehrels04}
{Gehrels}, N., {Chincarini}, G., {Giommi}, P., {et~al.} 2004, \apj, 611, 1005

\bibitem[{{Graham} {et~al.}(2020){Graham}, {Ross}, {Stern}, {Drake},
  {McKernan}, {Ford}, {Djorgovski}, {Mahabal}, {Glikman}, {Larson}, \&
  {Christensen}}]{graham20}
{Graham}, M.~J., {Ross}, N.~P., {Stern}, D., {et~al.} 2020, \mnras, 491, 4925

\bibitem[{{Grupe} {et~al.}(2019){Grupe}, {Komossa}, {Schartel}, {Parker},
  {Kollatschny}, {Ochmann}, \& {Colmenero}}]{grupe19}
{Grupe}, D., {Komossa}, S., {Schartel}, N., {et~al.} 2019, The Astronomer's
  Telegram, 13182, 1

\bibitem[{{Grupe} {et~al.}(2024){Grupe}, {Komossa}, \& {Wolsing}}]{grupe24}
{Grupe}, D., {Komossa}, S., \& {Wolsing}, S. 2024, \apj, 969, 98

\bibitem[{{Grupe} {et~al.}(2001){Grupe}, {Thomas}, \& {Beuermann}}]{grupe01}
{Grupe}, D., {Thomas}, H.~C., \& {Beuermann}, K. 2001, \aap, 367, 470

\bibitem[{{Hamuy} {et~al.}(1994){Hamuy}, {Suntzeff}, {Walker}, {Gigoux}, \&
  {Phillips}}]{Hamuy94}
{Hamuy}, M., {Suntzeff}, N.~B.~an{Heathcote}, S.~R., {Walker}, A.~R., {Gigoux},
  P., \& {Phillips}, M.~M. 1994, \pasp, 106, 566

\bibitem[{{Henden} {et~al.}(2012){Henden}, {Levine}, {Terrell}, {Smith}, \&
  {Welch}}]{henden12}
{Henden}, A.~A., {Levine}, S.~E., {Terrell}, D., {Smith}, T.~C., \& {Welch}, D.
  2012, JAAVSO, 40, 430

\bibitem[{{Hill} {et~al.}(2004){Hill}, {Burrows}, {Nousek}, {Abbey}, {Ambrosi},
  {Br{\"a}uninger}, {Burkert}, {Campana}, {Cheruvu}, {Cusumano}, {Freyberg},
  {Hartner}, {Klar}, {Mangels}, {Moretti}, {Mori}, {Morris}, {Short},
  {Tagliaferri}, {Watson}, {Wood}, \& {Wells}}]{hills05}
{Hill}, J.~E., {Burrows}, D.~N., {Nousek}, J.~A., {et~al.} 2004, in Society of
  Photo-Optical Instrumentation Engineers (SPIE) Conference Series, Vol. 5165,
  X-Ray and Gamma-Ray Instrumentation for Astronomy XIII, ed. K.~A. {Flanagan}
  \& O.~H.~W. {Siegmund}, 217--231

\bibitem[{{Husemann} {et~al.}(2016){Husemann}, {Urrutia}, {Tremblay}, {Krumpe},
  {Dexter}, {Busch}, {Combes}, {Croom}, {Davis}, {Eckart}, {McElroy},
  {Perez-Torres}, {Powell}, \& {Scharw{\"a}chter}}]{husemann16}
{Husemann}, B., {Urrutia}, T., {Tremblay}, G.~R., {et~al.} 2016, \aap, 593, L9

\bibitem[{{Jana} {et~al.}(2025){Jana}, {Ricci}, {Temple}, {Chang},
  {Shablovinskaya}, {Trakhtenbrot}, {Diaz}, {Ilic}, {Nandi}, \&
  {Koss}}]{jana25}
{Jana}, A., {Ricci}, C., {Temple}, M.~J., {et~al.} 2025, \aap, 693, A35

\bibitem[{{Jayasinghe} {et~al.}(2018){Jayasinghe}, {Kochanek}, {Stanek},
  {Shappee}, {Holoien}, {Thompson}, {Prieto}, {Dong}, {Pawlak}, {Shields},
  {Pojmanski}, {Otero}, {Britt}, \& {Will}}]{jayasinghe18}
{Jayasinghe}, T., {Kochanek}, C.~S., {Stanek}, K.~Z., {et~al.} 2018, \mnras,
  477, 3145

\bibitem[{{Kaaz} {et~al.}(2023){Kaaz}, {Liska}, {Jacquemin-Ide}, {Andalman},
  {Musoke}, {Tchekhovskoy}, \& {Porth}}]{kaaz23}
{Kaaz}, N., {Liska}, M. T.~P., {Jacquemin-Ide}, J., {et~al.} 2023, \apj, 955,
  72

\bibitem[{{Kim} {et~al.}(2018){Kim}, {Yoon}, \& {Evans}}]{kim18}
{Kim}, D.~C., {Yoon}, I., \& {Evans}, A.~S. 2018, \apj, 861, 51

\bibitem[{{Kochanek} {et~al.}(2017){Kochanek}, {Shappee}, {Stanek}, {Holoien},
  {Thompson}, {Prieto}, {Dong}, {Shields}, {Will}, {Britt}, {Perzanowski}, \&
  {Pojma{\'n}ski}}]{kochanek17}
{Kochanek}, C.~S., {Shappee}, B.~J., {Stanek}, K.~Z., {et~al.} 2017, \pasp,
  129, 104502

\bibitem[{{Kollatschny} {et~al.}(2000){Kollatschny}, {Bischoff}, \&
  {Dietrich}}]{kollatschny00}
{Kollatschny}, W., {Bischoff}, K., \& {Dietrich}, M. 2000, \aap, 361, 901

\bibitem[{{Kollatschny} {et~al.}(2001){Kollatschny}, {Bischoff}, {Robinson},
  {Welsh}, \& {Hill}}]{kollatschny01}
{Kollatschny}, W., {Bischoff}, K., {Robinson}, E.~L., {Welsh}, W.~F., \&
  {Hill}, G.~J. 2001, \aap, 379, 125

\bibitem[{{Kollatschny} \& {Chelouche}(2024)}]{kollatschny24}
{Kollatschny}, W. \& {Chelouche}, D. 2024, \aap, 690, L2

\bibitem[{{Kollatschny} \& {Fricke}(1985)}]{kollatschny85}
{Kollatschny}, W. \& {Fricke}, K.~J. 1985, \aap, 146, L11

\bibitem[{{Kollatschny} {et~al.}(2020){Kollatschny}, {Grupe}, {Parker},
  {Ochmann}, {Schartel}, {Herwig}, {Komossa}, {Romero-Colmenero}, \&
  {Santos-Lleo}}]{kollatschny20}
{Kollatschny}, W., {Grupe}, D., {Parker}, M.~L., {et~al.} 2020, \aap, 638, A91

\bibitem[{{Kollatschny} {et~al.}(2023){Kollatschny}, {Grupe}, {Parker},
  {Ochmann}, {Schartel}, {Romero-Colmenero}, {Winkler}, {Komossa}, {Famula},
  {Probst}, \& {Santos-Lleo}}]{kollatschny23}
{Kollatschny}, W., {Grupe}, D., {Parker}, M.~L., {et~al.} 2023, \aap, 670, A103

\bibitem[{{Kollatschny} {et~al.}(2022){Kollatschny}, {Ochmann}, {Kaspi},
  {Schumacher}, {Behar}, {Chelouche}, {Horne}, {M{\"u}ller}, {Rafter}, {Chini},
  {Haas}, \& {Probst}}]{kollatschny22}
{Kollatschny}, W., {Ochmann}, M.~W., {Kaspi}, S., {et~al.} 2022, \aap, 657,
  A122

\bibitem[{{Kollatschny} {et~al.}(2018){Kollatschny}, {Ochmann}, {Zetzl},
  {Haas}, {Chelouche}, {Kaspi}, {Pozo Nu{\~n}ez}, \& {Grupe}}]{kollatschny18}
{Kollatschny}, W., {Ochmann}, M.~W., {Zetzl}, M., {et~al.} 2018, \aap, 619,
  A168

\bibitem[{{Kollatschny} {et~al.}(2006){Kollatschny}, {Zetzl}, \&
  {Dietrich}}]{kollatschny06}
{Kollatschny}, W., {Zetzl}, M., \& {Dietrich}, M. 2006, \aap, 454, 459

\bibitem[{{Komossa}(2015)}]{komossa15}
{Komossa}, S. 2015, Journal of High Energy Astrophysics, 7, 148

\bibitem[{{Komossa} {et~al.}(2024){Komossa}, {Grupe}, {Marziani}, {Popovic},
  {Marceta-Mandic}, {Bon}, {Ilic}, {Kovacevic}, {Kraus}, {Haiman}, {Petrecca},
  {De Cicco}, {Dimitrijevic}, {Sreckovic}, {Kovacevic Dojcinovic},
  {Pannikkote}, {Bon}, {Gupta}, \& {Iacob}}]{komossa24}
{Komossa}, S., {Grupe}, D., {Marziani}, P., {et~al.} 2024, arXiv e-prints,
  arXiv:2408.00089

\bibitem[{{Komossa} {et~al.}(2008){Komossa}, {Zhou}, {Wang}, {Ajello}, {Ge},
  {Greiner}, {Lu}, {Salvato}, {Saxton}, {Shan}, {Xu}, \& {Yuan}}]{komossa08}
{Komossa}, S., {Zhou}, H., {Wang}, T., {et~al.} 2008, \apjl, 678, L13

\bibitem[{{Korista} \& {Goad}(2004)}]{korista04}
{Korista}, K.~T. \& {Goad}, M.~R. 2004, \apj, 606, 749

\bibitem[{{Laha} {et~al.}(2022){Laha}, {Meyer}, {Roychowdhury}, {Becerra
  Gonzalez}, {Acosta-Pulido}, {Thapa}, {Ghosh}, {Behar}, {Gallo}, {Kriss},
  {Panessa}, {Bianchi}, {La Franca}, {Scepi}, {Begelman}, {Longinotti},
  {Lusso}, {Oates}, {Nicholl}, \& {Cenko}}]{laha22}
{Laha}, S., {Meyer}, E., {Roychowdhury}, A., {et~al.} 2022, \apj, 931, 5

\bibitem[{{LaMassa} {et~al.}(2015){LaMassa}, {Cales}, {Moran}, {Myers},
  {Richards}, {Eracleous}, {Heckman}, {Gallo}, \& {Urry}}]{lamassa15}
{LaMassa}, S.~M., {Cales}, S., {Moran}, E.~C., {et~al.} 2015, \apj, 800, 144

\bibitem[{{Loveday}(1996)}]{loveday96}
{Loveday}, J. 1996, \mnras, 278, 1025

\bibitem[{{Lu} {et~al.}(2025){Lu}, {Li}, {Wu}, {Ho}, {Zhang}, {Feng}, {Li},
  {Chen}, {Sun}, {Shu}, {Guo}, {Cheng}, {Wang}, {Kim}, {Wang}, \& {Bai}}]{lu25}
{Lu}, K.-X., {Li}, Y.-R., {Wu}, Q., {et~al.} 2025, \apjs, 276, 51

\bibitem[{{Lusso} {et~al.}(2025){Lusso}, {Casetti}, {Romoli}, {Fossi},
  {Nardini}, {Arra}, {Barsi}, {Calamai}, {Campani}, {Capogrosso}, {Chiti
  Tegli}, {Ciantini}, {Demertzi}, {Gaitani}, {Giudice}, {Gori}, {Graziani},
  {Macchiarini}, {Michelagnoli}, {Niccolai}, {Parenti}, {Pistolesi}, {Rago},
  {Romani}, {Sani}, {Sartini}, {Scianni}, {Triggianese}, {Andreuzzi}, \&
  {Ambrosino}}]{lusso25}
{Lusso}, E., {Casetti}, L., {Romoli}, M., {et~al.} 2025, \aap, 695, A269

\bibitem[{{Lyu} {et~al.}(2021){Lyu}, {Yan}, {Yu}, \& {Wu}}]{lyu21}
{Lyu}, B., {Yan}, Z., {Yu}, W., \& {Wu}, Q. 2021, \mnras, 506, 4188

\bibitem[{{MacLeod} {et~al.}(2019){MacLeod}, {Green}, {Anderson}, {Bruce},
  {Eracleous}, {Graham}, {Homan}, {Lawrence}, {LeBleu}, {Ross}, {Ruan},
  {Runnoe}, {Stern}, {Burgett}, {Chambers}, {Kaiser}, {Magnier}, \&
  {Metcalfe}}]{macleod19}
{MacLeod}, C.~L., {Green}, P.~J., {Anderson}, S.~F., {et~al.} 2019, \apj, 874,
  8

\bibitem[{{McElroy} {et~al.}(2016){McElroy}, {Husemann}, {Croom}, {Davis},
  {Bennert}, {Busch}, {Combes}, {Eckart}, {Perez-Torres}, {Powell},
  {Scharw{\"a}chter}, {Tremblay}, \& {Urrutia}}]{mcelroy16}
{McElroy}, R.~E., {Husemann}, B., {Croom}, S.~M., {et~al.} 2016, \aap, 593, L8

\bibitem[{{Mehdipour} {et~al.}(2022){Mehdipour}, {Kriss}, {Brenneman},
  {Costantini}, {Kaastra}, {Branduardi-Raymont}, {Di Gesu}, {Ebrero}, \&
  {Mao}}]{mehdipour22}
{Mehdipour}, M., {Kriss}, G.~A., {Brenneman}, L.~W., {et~al.} 2022, \apj, 925,
  84

\bibitem[{{Moshir} \& {et al.}(1990)}]{moshir90}
{Moshir}, M. \& {et al.} 1990, IRAS Faint Source Catalogue, 0

\bibitem[{{Netzer}(2019)}]{netzer19}
{Netzer}, H. 2019, \mnras, 488, 5185

\bibitem[{{Ochmann} {et~al.}(2024){Ochmann}, {Kollatschny}, {Probst},
  {Romero-Colmenero}, {Buckley}, {Chelouche}, {Chini}, {Grupe}, {Haas},
  {Kaspi}, {Komossa}, {Parker}, {Santos-Lleo}, {Schartel}, \&
  {Famula}}]{ochmann24}
{Ochmann}, M.~W., {Kollatschny}, W., {Probst}, M.~A., {et~al.} 2024, \aap, 686,
  A17

\bibitem[{{Oknyansky} {et~al.}(2019){Oknyansky}, {Winkler}, {Tsygankov},
  {Lipunov}, {Gorbovskoy}, {van Wyk}, {Buckley}, \& {Tyurina}}]{Oknyansky19}
{Oknyansky}, V.~L., {Winkler}, H., {Tsygankov}, S.~S., {et~al.} 2019, \mnras,
  483, 558

\bibitem[{{Panda} \& {{\'S}niegowska}(2024)}]{panda24}
{Panda}, S. \& {{\'S}niegowska}, M. 2024, \apjs, 272, 13

\bibitem[{{Park} {et~al.}(2006){Park}, {Kashyap}, {Siemiginowska}, {van Dyk},
  {Zezas}, {Heinke}, \& {Wargelin}}]{park06}
{Park}, T., {Kashyap}, V.~L., {Siemiginowska}, A., {et~al.} 2006, \apj, 652,
  610

\bibitem[{{Parker} {et~al.}(2016){Parker}, {Komossa}, {Kollatschny}, {Walton},
  {Schartel}, {Santos-Lle{\'o}}, {Harrison}, {Fabian}, {Zetzl}, {Grupe},
  {Rodr{\'\i}guez-Pascual}, \& {Vasudevan}}]{parker16}
{Parker}, M.~L., {Komossa}, S., {Kollatschny}, W., {et~al.} 2016, \mnras, 461,
  1927

\bibitem[{{Parker} {et~al.}(2019){Parker}, {Schartel}, {Grupe}, {Komossa},
  {Harrison}, {Kollatschny}, {Mikula}, {Santos-Lle{\'o}}, \&
  {Tom{\'a}s}}]{parker19}
{Parker}, M.~L., {Schartel}, N., {Grupe}, D., {et~al.} 2019, \mnras, 483, L88

\bibitem[{{Pastoriza} \& {Gerola}(1970)}]{pastroriza70}
{Pastoriza}, M. \& {Gerola}, H. 1970, \aplett, 6, 155

\bibitem[{{Penston} \& {Perez}(1984)}]{penston84}
{Penston}, M.~V. \& {Perez}, E. 1984, \mnras, 211, 33P

\bibitem[{{Poole} {et~al.}(2008){Poole}, {Breeveld}, {Page}, {Landsman},
  {Holland}, {Roming}, {Kuin}, {Brown}, {Gronwall}, {Hunsberger}, {Koch},
  {Mason}, {Schady}, {vanden Berk}, {Blustin}, {Boyd}, {Broos}, {Carter},
  {Chester}, {Cucchiara}, {Hancock}, {Huckle}, {Immler}, {Ivanushkina},
  {Kennedy}, {Marshall}, {Morgan}, {Pandey}, {de Pasquale}, {Smith}, \&
  {Still}}]{poole08}
{Poole}, T.~S., {Breeveld}, A.~A., {Page}, M.~J., {et~al.} 2008, \mnras, 383,
  627

\bibitem[{{Popovi{\'c}} {et~al.}(2023){Popovi{\'c}}, {Ili{\'c}}, {Burenkov},
  {Pati{\~n}o Alvarez}, {Mar{\v{c}}eta-Mandi{\'c}},
  {Kova{\v{c}}evi{\'c}-Doj{\v{c}}inovi{\'c}}, {Shablovinskaya},
  {Kova{\v{c}}evi{\'c}}, {Marziani}, {Chavushyan}, {Wang}, {Li}, \&
  {Mediavilla}}]{popovic23}
{Popovi{\'c}}, L.~{\v{C}}., {Ili{\'c}}, D., {Burenkov}, A., {et~al.} 2023,
  \aap, 675, A178

\bibitem[{{Ricci} {et~al.}(2020){Ricci}, {Kara}, {Loewenstein}, {Trakhtenbrot},
  {Arcavi}, {Remillard}, {Fabian}, {Gendreau}, {Arzoumanian}, {Li}, {Ho},
  {MacLeod}, {Cackett}, {Altamirano}, {Gandhi}, {Kosec}, {Pasham}, {Steiner},
  \& {Chan}}]{ricci20}
{Ricci}, C., {Kara}, E., {Loewenstein}, M., {et~al.} 2020, \apjl, 898, L1

\bibitem[{{Ricci} \& {Trakhtenbrot}(2023)}]{ricci23}
{Ricci}, C. \& {Trakhtenbrot}, B. 2023, Nature Astronomy, 7, 1282

\bibitem[{{Rodr{\'\i}guez-Pascual} {et~al.}(1997){Rodr{\'\i}guez-Pascual},
  {Alloin}, {Clavel}, {Crenshaw}, {Horne}, {Kriss}, {Krolik}, {Malkan},
  {Netzer}, {O'Brien}, {Peterson}, {Reichert}, {Wamsteker}, {Alexander},
  {Barr}, {Blandford}, {Bregman}, {Carone}, {Clements}, {Courvoisier}, {De
  Robertis}, {Dietrich}, {Dottori}, {Edelson}, {Filippenko}, {Gaskell},
  {Huchra}, {Hutchings}, {Kollatschny}, {Koratkar}, {Korista}, {Laor},
  {MacAlpine}, {Martin}, {Maoz}, {McCollum}, {Morris}, {Perola}, {Pogge},
  {Ptak}, {Recondo-Gonz{\'a}lez}, {Rodr{\'\i}guez-Espinoza}, {Rokaki},
  {Santos-Lle{\'o}}, {Sekiguchi}, {Shull}, {Snijders}, {Sparke}, {Stirpe},
  {Stoner}, {Sun}, {Wagner}, {Wanders}, {Wilkes}, {Winge}, \&
  {Zheng}}]{rodriguez97}
{Rodr{\'\i}guez-Pascual}, P.~M., {Alloin}, D., {Clavel}, J., {et~al.} 1997,
  \apjs, 110, 9

\bibitem[{{Roming} {et~al.}(2005){Roming}, {Kennedy}, {Mason}, {Nousek}, {Ahr},
  {Bingham}, {Broos}, {Carter}, {Hancock}, {Huckle}, {Hunsberger}, {Kawakami},
  {Killough}, {Koch}, {McLelland}, {Smith}, {Smith}, {Soto}, {Boyd},
  {Breeveld}, {Holland}, {Ivanushkina}, {Pryzby}, {Still}, \&
  {Stock}}]{roming05}
{Roming}, P. W.~A., {Kennedy}, T.~E., {Mason}, K.~O., {et~al.} 2005, \ssr, 120,
  95

\bibitem[{{Rumbaugh} {et~al.}(2018){Rumbaugh}, {Shen}, {Morganson}, {Liu},
  {Banerji}, {McMahon}, {Abdalla}, {Benoit-L{\'e}vy}, {Bertin}, {Brooks},
  {Buckley-Geer}, {Capozzi}, {Carnero Rosell}, {Carrasco Kind}, {Carretero},
  {Cunha}, {D'Andrea}, {da Costa}, {DePoy}, {Desai}, {Doel}, {Frieman},
  {Garc{\'\i}a-Bellido}, {Gruen}, {Gruendl}, {Gschwend}, {Gutierrez},
  {Honscheid}, {James}, {Kuehn}, {Kuhlmann}, {Kuropatkin}, {Lima}, {Maia},
  {Marshall}, {Martini}, {Menanteau}, {Plazas}, {Reil}, {Roodman}, {Sanchez},
  {Scarpine}, {Schindler}, {Schubnell}, {Sheldon}, {Smith}, {Soares-Santos},
  {Sobreira}, {Suchyta}, {Swanson}, {Walker}, {Wester}, \& {DES
  Collaboration}}]{rumbaugh18}
{Rumbaugh}, N., {Shen}, Y., {Morganson}, E., {et~al.} 2018, \apj, 854, 160

\bibitem[{{Schlafly} \& {Finkbeiner}(2011)}]{schlafly11}
{Schlafly}, E.~F. \& {Finkbeiner}, D.~P. 2011, \apj, 737, 103

\bibitem[{{Shappee} {et~al.}(2014){Shappee}, {Prieto}, {Grupe}, {Kochanek},
  {Stanek}, {De Rosa}, {Mathur}, {Zu}, {Peterson}, {Pogge}, {Komossa}, {Im},
  {Jencson}, {Holoien}, {Basu}, {Beacom}, {Szczygie{\l}}, {Brimacombe},
  {Adams}, {Campillay}, {Choi}, {Contreras}, {Dietrich}, {Dubberley},
  {Elphick}, {Foale}, {Giustini}, {Gonzalez}, {Hawkins}, {Howell}, {Hsiao},
  {Koss}, {Leighly}, {Morrell}, {Mudd}, {Mullins}, {Nugent}, {Parrent},
  {Phillips}, {Pojmanski}, {Rosing}, {Ross}, {Sand}, {Terndrup}, {Valenti},
  {Walker}, \& {Yoon}}]{shappee14}
{Shappee}, B.~J., {Prieto}, J.~L., {Grupe}, D., {et~al.} 2014, \apj, 788, 48

\bibitem[{{Shectman} {et~al.}(1996){Shectman}, {Landy}, {Oemler}, {Tucker},
  {Lin}, {Kirshner}, \& {Schechter}}]{shectman96}
{Shectman}, S.~A., {Landy}, S.~D., {Oemler}, A., {et~al.} 1996, \apj, 470, 172

\bibitem[{{Stern} {et~al.}(2018){Stern}, {McKernan}, {Graham}, {Ford}, {Ross},
  {Meisner}, {Assef}, {Balokovi{\'c}}, {Brightman}, {Dey}, {Drake},
  {Djorgovski}, {Eisenhardt}, \& {Jun}}]{stern18}
{Stern}, D., {McKernan}, B., {Graham}, M.~J., {et~al.} 2018, \apj, 864, 27

\bibitem[{{Storchi-Bergmann} {et~al.}(1997){Storchi-Bergmann}, {Eracleous},
  {Teresa Ruiz}, {Livio}, {Wilson}, \& {Filippenko}}]{storchi97}
{Storchi-Bergmann}, T., {Eracleous}, M., {Teresa Ruiz}, M., {et~al.} 1997,
  \apj, 489, 87

\bibitem[{{Tohline} \& {Osterbrock}(1976)}]{tohline76}
{Tohline}, J.~E. \& {Osterbrock}, D.~E. 1976, \apjl, 210, L117

\bibitem[{{Trakhtenbrot} {et~al.}(2019){Trakhtenbrot}, {Arcavi}, {MacLeod},
  {Ricci}, {Kara}, {Graham}, {Stern}, {Harrison}, {Burke}, {Hiramatsu},
  {Hosseinzadeh}, {Howell}, {Smartt}, {Rest}, {Prieto}, {Shappee}, {Holoien},
  {Bersier}, {Filippenko}, {Brink}, {Zheng}, {Li}, {Remillard}, \&
  {Loewenstein}}]{trakhtenbrot19}
{Trakhtenbrot}, B., {Arcavi}, I., {MacLeod}, C.~L., {et~al.} 2019, \apj, 883,
  94

\bibitem[{{Trakhtenbrot} \& {Netzer}(2012)}]{trakhtenbrot12}
{Trakhtenbrot}, B. \& {Netzer}, H. 2012, \mnras, 427, 3081

\bibitem[{{van Velzen} {et~al.}(2021){van Velzen}, {Gezari}, {Hammerstein},
  {Roth}, {Frederick}, {Ward}, {Hung}, {Cenko}, {Stein}, {Perley}, {Taggart},
  {Foley}, {Sollerman}, {Blagorodnova}, {Andreoni}, {Bellm}, {Brinnel}, {De},
  {Dekany}, {Feeney}, {Fremling}, {Giomi}, {Golkhou}, {Graham}, {Ho},
  {Kasliwal}, {Kilpatrick}, {Kulkarni}, {Kupfer}, {Laher}, {Mahabal}, {Masci},
  {Miller}, {Nordin}, {Riddle}, {Rusholme}, {van Santen}, {Sharma}, {Shupe}, \&
  {Soumagnac}}]{vanvelzen21}
{van Velzen}, S., {Gezari}, S., {Hammerstein}, E., {et~al.} 2021, \apj, 908, 4

\bibitem[{von Neumann(1941)}]{neumann41}
von Neumann, J. 1941, The Annals of Mathematical Statistics, 12, 367

\bibitem[{{Wang} {et~al.}(2024){Wang}, {Xu}, {Cao}, {Gao}, {Xie}, \&
  {Wei}}]{wang24}
{Wang}, J., {Xu}, D.~W., {Cao}, X., {et~al.} 2024, \apj, 970, 85

\bibitem[{{Wang} {et~al.}(2025){Wang}, {Woo}, {Gallo}, {Son}, {Yang}, {Jin},
  {Guo}, \& {Kong}}]{wang25}
{Wang}, S., {Woo}, J.-H., {Gallo}, E., {et~al.} 2025, \apj, 981, 129

\bibitem[{{Ward} {et~al.}(2025){Ward}, {Koss}, {Eracleous}, {Trakhtenbrot},
  {Bauer}, {Caglar}, {Harrison}, {Jana}, {Kakkad}, {Magno}, {del Moral-Castro},
  {Mushotzky}, {Oh}, {Peca}, {Powell}, {Ricci}, {Rojas}, {Smith}, {Stern},
  {Treister}, \& {Urry}}]{ward25}
{Ward}, C., {Koss}, M.~J., {Eracleous}, M., {et~al.} 2025, arXiv e-prints,
  arXiv:2507.05380

\bibitem[{{Wright}(2006)}]{wright06}
{Wright}, E.~L. 2006, \pasp, 118, 1711

\bibitem[{{Yin} {et~al.}(2025){Yin}, {Lawrence}, {Ward}, {Homan}, \&
  {Kollatschny}}]{yin25}
{Yin}, C., {Lawrence}, A., {Ward}, M., {Homan}, D., \& {Kollatschny}, W. 2025,
  \mnras, 540, 3032

\bibitem[{{Zeltyn} {et~al.}(2024){Zeltyn}, {Trakhtenbrot}, {Eracleous}, {Yang},
  {Green}, {Anderson}, {LaMassa}, {Runnoe}, {Assef}, {Bauer}, {Brandt},
  {Davis}, {Frederick}, {Fries}, {Graham}, {Grogin}, {Guolo},
  {Hern{\'a}ndez-Garc{\'\i}a}, {Koekemoer}, {Krumpe}, {Liu},
  {Mart{\'\i}nez-Aldama}, {Ricci}, {Schneider}, {Shen}, {{\'S}niegowska},
  {Temple}, {Trump}, {Xue}, {Brownstein}, {Dwelly}, {Morrison}, {Bizyaev},
  {Pan}, \& {Kollmeier}}]{zeltyn24}
{Zeltyn}, G., {Trakhtenbrot}, B., {Eracleous}, M., {et~al.} 2024, \apj, 966, 85

\bibitem[{{Zetzl} {et~al.}(2018){Zetzl}, {Kollatschny}, {Ochmann}, {Grupe},
  {Haas}, {Ramolla}, {Chelouche}, {Kaspi}, \& {Schartel}}]{zetzl18}
{Zetzl}, M., {Kollatschny}, W., {Ochmann}, M.~W., {et~al.} 2018, \aap, 618, A83

\end{thebibliography}

\onecolumn

\begin{appendix}
\section{\bf Additional tables and figures}

\begin{figure}[!h]
\centering
\includegraphics[width=17.cm,angle=0]{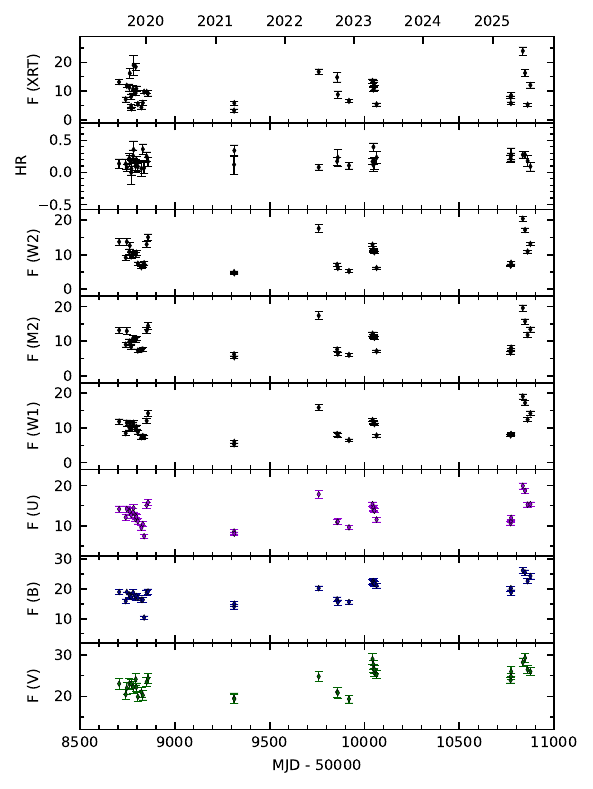}
\caption{Combined X-ray, UV, and optical light curves taken with the \swift{} satellite for the years 2019 to 2025. The fluxes are given in units of 10$^{-12}$ ergs s$^{-1}$ cm$^{-2}$. HR is the X-ray hardness ratio. }
\label{DirkLcIras23226zoom_xrt_uvot_lc_20250819}
\end{figure}

\newpage

\begin{table}[!t]
\caption
{  {\swift{} monitoring: 
MJD, UT date, XRT 0.3--10 keV count rates (CR),
hardness ratios (HR$^1$), X-ray photon index $\Gamma$, the absorption corrected 
0.3--10 keV X-ray flux in units of $10^{-12}$ erg s$^{-1}$ cm$^{-2}$, 
Cash Statistic (CSTAT), and degree of freedom (dof).}
}
\small
\tabcolsep+1mm
\begin{tabular*}{\textwidth}{@{\extracolsep{\fill} } lcccccrr}
\hline 
\noalign{\smallskip}
& UT date & \\
\rb{MJD} & middle of the exposure & \rb{CR} & \rb{HR$^1$} &  \rb{$\Gamma_{\rm pl}$} & \rb{XRT Flux} & \rb{CSTAT} & \rb{dof}\\
\hline 
54266.1750  &2007-06-15 04:12  & 0.286\plm0.011  & +0.208\plm0.024 & 1.76\plm0.07 & 11.54\plm0.47 &  363.17 & 373 \\
54312.4792  &2007-07-31 11:30 &  0.238\plm0.008  & +0.178\plm0.028 & 1.85\plm0.07 &  8.26\plm0.30 &  276.78 & 341\\
54362.5389  &2007-09-19 12:56 & 0.086\plm0.004  & +0.386\plm0.051  & 1.51\plm0.12 &  3.80\plm0.33 &  172.71 & 234 \\
54363.6139  &2007-09-20 14:44 & 0.079\plm0.003  & +0.255\plm0.044  & 1.70\plm0.11 &  3.15\plm0.18 &  240.75 & 246\\
54676.2875  &2008-07-29 06:54 & 0.288\plm0.020  & +0.311\plm0.055  & 1.57\plm0.16 & 12.71\plm1.25 &  136.61 & 161\\
55092.4722  &2009-09-18 11:20 & 0.340\plm0.007  & +0.138\plm0.015  & 1.81\plm0.04 & 11.82\plm0.28 &  535.02 & 541 \\
56492.4166  &2013-07-20 10:13 & 0.073\plm0.008  & +0.305\plm0.096  & 1.55\plm0.27 &  3.42\plm0.30 &   65.79 & 74\\
56555.8861  &2013-09-20 21:16 &0.149\plm0.018  & +0.241\plm0.127   & 1.64\plm0.34 &  6.40\plm1.02 &   50.82 & 53 \\
57577.5396  &2016-07-08 12:57 &  0.051\plm0.003  & +0.345\plm0.056 & 1.61\plm0.16 &  2.22\plm0.34 &  161.54 & 151\\
57855.7061  &2017-04-12 16:56 & 0.037\plm0.009  & +0.212\plm0.240  & ----         &  ----         &   12.77 & 13\\
57855.8778  &2017-04-12 21:04 & 0.045\plm0.010  & +0.367\plm0.210  & ----         &  ----         &   18.92 & 16\\
57864.5097  &2017-04-21 12:14 & 0.045\plm0.010  & +0.057\plm0.177  & ----         &  ----         &   19.66 & 23\\
57880.2952  &2017-05-07 07:05 & 0.040\plm0.008  & +0.562\plm0.136  & ----         &  ----         &  ----   & 27 \\
57891.1903  &2017-05-18 04:34 & 0.029\plm0.010  & +0.383\plm0.232  & ---          &  ----         &  ----   & 10 \\
57894.6028  &2017-05-21 14:28 & 0.032\plm0.008  & +0.554\plm0.153  & ----         & ----          &   16.55 & 23\\
57895.0243  &2017-05-23 00:35 & 0.024\plm0.006  & -0.224\plm0.241  & ----         & ----          &   13.85 & 14  \\
57902.3451  &2017-05-29 08:17 &  0.018\plm0.007  & +0.348\plm0.310 & ---          & -----         &    9.25 & 4\\
57912.7708  &2017-06-08 18:30 & 0.030\plm0.007  & -0.002\plm0.220  & ----         & ----          &   12.48 & 15 \\
57915.9201  &2017-06-11 22:06 & 0.018\plm0.004  & +0.028\plm0.200  & ----         & ----          &   18.40 & 26\\
57922.9986  &2017-06-18 23:59 & 0.016\plm0.005  & -0.211\plm0.310  &  ---         & ----          &    6.61 & 10\\
57930.7806  &2017-06-26 18:44 & 0.014\plm0.007  & ----             & ---          & ----          &    6.61 & 10\\
57937.6931  &2017-07-03 16:37 & 0.021\plm0.006  & +0.056\plm0.190  & ---          & ----          &    4.65 & 8\\
57944.0639  &2017-07-10 01:32 & 0.022\plm0.006  & +0.435\plm0.221  & ---          & ----          &    8.69 & 11\\
58705.4279  &2019-08-11 10:16 & 0.403\plm0.033  & +0.137\plm0.074  & 1.89\plm0.16 & 13.20\plm0.81 &  109.19 & 148\\
58740.6404  &2019-09-14 15:22 & 0.200\plm0.018  & +0.133\plm0.079  & 1.94\plm0.23 &  7.23\plm0.87 &   81.11 & 93\\
58746.2726  &2019-09-20 06:32 & 0.355\plm0.042  & +0.073\plm0.053  & 2.05\plm0.22 & 11.88\plm0.60 &   59.61 & 96\\
58759.5462  &2019-10-03 13:06 & 0.298\plm0.030  & +0.219\plm0.073  & 1.83\plm0.17 & 11.34\plm1.20 &  123.86 & 135\\
58761.9466  &2019-10-05 22:43 & 0.400\plm0.048  & +0.186\plm0.083  & 1.96\plm0.21 & 16.21\plm1.66 &   66.44 & 102\\
58769.3119  &2019-10-13 07:29 & 0.161\plm0.014  & +0.160\plm0.058  & 1.95\plm0.25 &  4.86\plm0.63 &   62.77 & 82\\
58769.4993  &2019-10-13 11:59 & 0.179\plm0.014  & +0.000\plm0.190  & 1.71\plm0.20 &  8.16\plm0.95 &   86.33 & 116\\
58770.7026  &2019-10-14 16:51 & 0.148\plm0.013  & +0.056\plm0.094  & 2.04\plm0.25 &  4.05\plm0.50 &   64.34 & 78\\
58779.6649  &2019-10-23 15:57 & 0.303\plm0.028  & +0.257\plm0.093  & 1.96\plm0.24 & 10.74\plm1.33 &   75.31 & 84\\
58782.0028  &2019-10-26 00:04 & 0.337\plm0.049  & +0.355\plm0.135  & 1.41\plm0.44 & 19.04\plm3.43 &   26.61 & 35\\
58792.7181  &2019-11-05 17:14 & 0.265\plm0.016  & +0.095\plm0.064  & 1.90\plm0.14 &  9.37\plm0.85 &  157.88 & 179\\
58793.4931  &2019-11-06 11:50 & 0.266\plm0.016  & +0.108\plm0.099  & 1.82\plm0.24 & 18.47\plm1.29 &   64.24 & 81\\
58796.5980  &2019-11-09 14:21 &  0.306\plm0.022  & +0.198\plm0.049 & 1.73\plm0.16 & 10.84\plm0.99 &  120.29 & 145\\
58804.5670  &2019-11-17 13:36 & 0.154\plm0.015  & +0.091\plm0.076  & 1.98\plm0.25 &  5.59\plm0.58 &   50.00 & 75\\
58823.3669  &2019-12-06 08:48 & 0.127\plm0.015  & +0.062\plm0.125  & 1.84\plm0.30 &  4.46\plm0.59 &   40.26 & 60\\
58830.8669  &2019-12-13 20:48 & 0.152\plm0.014  & +0.362\plm0.081  & 1.67\plm0.22 &  5.99\plm0.79 &   79.51 & 87\\
58837.5791  &2019-12-20 13:53 & 0.291\plm0.033  & +0.070\plm0.087  & 1.87\plm0.23 &  9.82\plm0.49 &   58.68 & 83\\
58851.0457  &2020-01-03 01:05 & 0.233\plm0.018  & +0.247\plm0.072  & 1.71\plm0.17 &  9.78\plm0.52 &  104.34 & 142\\
58858.2817  &2020-01-10 06:45 & 0.308\plm0.023  & +0.168\plm0.071  & 1.92\plm0.17 &  9.32\plm0.96 &  114.75 & 151\\
59312.1285  &2021-04-08 03:05 & 0.085\plm0.013  & +0.121\plm0.150  & 1.95\plm0.40 &  3.30\plm0.65 &   33.05 & 32\\
59313.2613  &2021-04-09 06:16 & 0.138\plm0.014  & +0.341\plm0.088  & 1.61\plm0.25 &  5.89\plm0.67 &   63.70 & 76\\
59758.5143  &2022-06-28 12:20 & 0.567\plm0.033  & +0.081\plm0.041  & 1.94\plm0.11 & 16.74\plm1.01 &  213.27 & 238 \\
59855.6494  &2022-10-03 15:35 & 0.412\plm0.035  & +0.173\plm0.067  & 1.73\plm0.16 & 14.83\plm1.62 &  115.57 & 158\\
59859.3600  &2022-10-07 08:38 & 0.228\plm0.032  & +0.231\plm0.131  & 1.78\plm0.35 &  8.86\plm1.26 &   48.10 & 41\\
59918.0988  &2022-12-05 02:22 & 0.161\plm0.010  & +0.108\plm0.054  & 1.77\plm0.14 &  6.65\plm0.45 &  139.92 & 142 \\
60042.2293  &2023-04-08 05:30 & 0.397\plm0.023  & +0.169\plm0.047  & 1.82\plm0.12 & 13.56\plm0.72 &  190.73 & 225 \\
60046.5968  &2023-04-12 14:19 & 0.275\plm0.015  & +0.117\plm0.101  & 1.84\plm0.12 & 10.61\plm0.66 &  203.92 & 212 \\
60047.2907  &2023-04-13 06:58 & 0.265\plm0.018  & +0.397\plm0.065  & 1.53\plm0.17 & 12.84\plm1.09 &  123.22 & 147 \\
60051.1603  &2023-04-17 03:50 & 0.343\plm0.019  & +0.184\plm0.051  & 1.84\plm0.13 & 12.66\plm0.60 &  159.68 & 215 \\
60052.3498  &2023-04-18 08:23 & 0.287\plm0.015  & +0.145\plm0.094  & 1.86\plm0.12 & 11.22\plm0.77 &  185.55 & 214 \\
60063.6655  &2023-04-29 15:58 & 0.142\plm0.012  & +0.232\plm0.092  & 1.74\plm0.21 &  5.45\plm0.49 &   90.23 & 102 \\
60770.7224  &2025-04-05 17:20 & 0.205\plm0.011  & +0.219\plm0.056  & 1.70\plm0.14 &  7.92\plm0.48 &  159.37 & 198 \\
60772.1852  &2025-04-07 04:26 & 0.149\plm0.007  & +0.253\plm0.052  & 1.75\plm0.12 &  5.88\plm0.46 &  168.79 & 234 \\
60773.2583  &2025-04-08 06:11 & 0.218\plm0.020  & +0.282\plm0.102  & 1.71\plm0.23 &  8.61\plm0.96 &   72.17 & 90\\
60834.9438  &2025-06-08 22:39 & 0.737\plm0.047  & +0.275\plm0.049  & 1.77\plm0.12 & 23.96\plm1.45 &  217.09 & 228 \\
60846.8062  &2025-06-20 19:20 & 0.404\plm0.029  & +0.274\plm0.053  & 1.66\plm0.13 & 16.31\plm1.27 &  180.34 & 204 \\
60860.2544  &2025-07-04 06:06 & 0.136\plm0.012  & +0.179\plm0.089  & 1.78\plm0.23 &  5.33\plm0.58 &   77.74 & 94 \\
60874.9277  &2025-07-18 22:15 & 0.396\plm0.033  & +0.093\plm0.067  & 2.02\plm0.17 & 12.13\plm0.98 &  119.12 & 144 \\
\hline 
\end{tabular*}
$^1$ The hardness ratio is defined as HR = $\frac{hard-soft}{hard+soft}$ , where {\it \textup{soft}} and {\it \textup{hard}} are the background-corrected counts in the 0.3--1.0 keV and 1.0--10.0 keV bands, respectively 

\label{swiftdata} 
\end{table}

\begin{table}[!t]
\centering
\tabcolsep1.mm
\caption{  {\swift{} monitoring: V, B, U, UVOT W1, M2, and W2 reddening-corrected flux  in units of 10$^{-15}$W\,m$^{-2}$(10$^{-12}$ ergs s$^{-1}$ cm$^{-2}$).}
}
\small
\begin{tabular}{lcccccc}
\hline
\noalign{\smallskip}
MJD & V & B & U &UVW1  & UVM2 & UVW2  \\
\noalign{\smallskip}
(1) & (2) & (3) & (4) & (5) & (6) & (7) 
 \\  
\noalign{\smallskip}
\hline
\noalign{\smallskip}
54266.1750 &   ---          &   ---          &     ---      &    ---    &     --- &   9.85\plm0.14   \\
54312.4792 &   ---          &   ---          &   11.16\plm0.52 &    --- &     --- &     ---        \\
54362.5389 &   ---          &   ---          &   ---     &    ---       &     --- &     ---        \\
54363.6139 &   ---          &   ---          &   ---     &    ---       &     --- &     ---         \\
54676.2875 &  22.44\plm1.05 &  17.32\plm0.81 &   11.80\plm0.55 &   9.57\plm0.54  &    9.92\plm0.65  &   8.42\plm0.55   \\       
55092.4722 &    ---         &   ---          &   ---           &    ---          &      ---         &     ---         \\   
56492.9166 &  17.50\plm0.82 &  12.66\plm0.59 &    6.79\plm0.38 &   4.62\plm0.30  &    3.47\plm0.23  &   3.47\plm0.23   \\   
56555.8861 &  19.73\plm1.30 &  14.40\plm0.81 &    7.72\plm0.51 &   5.61\plm0.42  &    4.70\plm0.45  &   4.58\plm0.39   \\  
57577.5396 &  17.02\plm1.12 &   4.68\plm0.31 &    5.75\plm0.38 &   3.63\plm0.27  &    2.88\plm0.24  &   2.68\plm0.15   \\   
57855.2061 &  17.82\plm1.18 &  11.23\plm0.74 &    5.24\plm0.39 &   2.71\plm0.28  &    2.70\plm0.31  &   2.49\plm0.23   \\
57855.5423 &  17.50\plm1.15 &  12.09\plm0.80 &    5.05\plm0.38 &   2.97\plm0.31  &    2.65\plm0.30  &   2.73\plm0.26   \\   
57864.5097 &  16.26\plm0.91 &  11.76\plm0.55 &    5.10\plm0.33 &   3.19\plm0.24  &    2.58\plm0.22  &   2.78\plm0.21   \\   
57880.2952 &  17.18\plm0.97 &  11.98\plm0.56 &    5.91\plm0.39 &   2.99\plm0.22  &    2.47\plm0.23  &   2.61\plm0.22   \\  
57891.1903 &  17.34\plm1.14 &  12.89\plm0.72 &    5.15\plm0.39 &   3.14\plm0.30  &    2.53\plm0.26  &   2.59\plm0.24   \\  
57894.6028 &  17.82\plm1.00 &  10.82\plm0.61 &    5.19\plm0.34 &   2.31\plm0.22  &    2.51\plm0.24  &   2.63\plm0.22   \\   
57895.5243 &  16.71\plm0.94 &  11.87\plm0.55 &    5.64\plm0.37 &   2.99\plm0.22  &    2.33\plm0.22  &   2.47\plm0.21   \\   
57902.3451 &  17.82\plm1.00 &  11.76\plm0.55 &    5.19\plm0.34 &   2.86\plm0.24  &    2.86\plm0.24  &   2.29\plm0.19  \\  
57912.7708 &  18.16\plm1.02 &  11.87\plm0.55 &    4.78\plm0.31 &   2.78\plm0.23  &    2.49\plm0.23  &   2.47\plm0.21   \\  
57915.9201 &  15.67\plm0.73 &  11.23\plm0.52 &    5.29\plm0.29 &   2.89\plm0.19  &    2.44\plm0.18  &   2.38\plm0.18   \\  
57922.9986 &  17.50\plm0.98 &  11.65\plm0.54 &    5.24\plm0.34 &   2.86\plm0.21  &    2.49\plm0.23  &   2.49\plm0.21   \\  
57930.7806 &  16.41\plm1.24 &  12.09\plm0.80 &    5.49\plm0.47 &   2.89\plm0.30  &    2.40\plm0.35  &   2.54\plm0.26   \\  
57937.6931 &  17.99\plm1.01 &  12.66\plm0.71 &    4.69\plm0.35 &   2.94\plm0.25  &    2.19\plm0.23  &   2.34\plm0.20   \\  
57944.0639 &  16.56\plm0.93 &  12.20\plm0.57 &    4.78\plm0.31 &   2.89\plm0.21  &    1.98\plm0.37  &   2.68\plm0.20   \\  
58705.9279 &  23.07\plm1.30 &  18.99\plm0.89 &   14.18\plm0.80 &  11.83\plm0.78  &   13.20\plm0.87  &  13.72\plm1.04   \\  
58740.6404 &  20.47\plm1.15 &  15.94\plm0.74 &   12.13\plm0.68 &   8.49\plm0.56  &    9.05\plm0.68  &   9.15\plm0.69   \\  
58746.2726 &  22.03\plm1.45 &  18.99\plm0.10 &   14.18\plm0.80 &  11.50\plm0.87  &   12.96\plm0.98  &  13.72\plm1.04   \\  
58759.5462 &  23.07\plm1.30 &  18.13\plm0.85 &   13.80\plm0.77 &  10.11\plm0.67  &    9.92\plm0.75  &  10.90\plm0.72   \\ 
58761.9466 &    ---         &  17.64\plm0.82 &   13.54\plm0.76 &  11.29\plm0.74  &     ---          &  12.63\plm0.96   \\ 
58769.3119 &  23.07\plm1.08 &  17.32\plm0.81 &   12.70\plm0.71 &   9.84\plm0.65  &    9.92\plm0.65  &   9.49\plm0.62   \\ 
58769.5091 &    ---         &   ---          &   ---           &    ---          &    8.41\plm0.55  &      ---          \\
58770.7026 &    ---         &   ---          &   ---           &    ---          &     ---          &    9.76\plm0.83   \\ 
58779.6649 & 22.23\plm1.47  &  18.81\plm1.08 &   13.42\plm0.88  &  10.99\plm0.73  &   10.88\plm0.83  &  10.12\plm0.87   \\
58782.0028 &    ---         &   ---          &   14.45\plm0.95 &  11.50\plm0.65  &     ---          &     ---           \\  
58792.7181 &    ---         &   ---          &   11.69\plm0.43 &    ---          &     ---          &      ---           \\ 
58793.4931 &  24.16\plm1.36 &  17.16\plm0.96 &   12.47\plm0.70 &   9.75\plm0.64  &   10.58\plm0.70  &  10.03\plm0.66   \\ 
58796.5980 &  22.23\plm1.04 &  17.64\plm0.82 &   12.24\plm0.69 &  10.30\plm0.58  &   10.78\plm0.60  &  10.60\plm0.70   \\  
58804.5670 &  19.91\plm1.12 &  17.00\plm0.79 &   11.16\plm0.63 &   8.81\plm0.49  &    7.32\plm0.48  &   7.40\plm0.49   \\ 
58823.3669 &  21.04\plm1.18 &  16.38\plm0.76 &    9.54\plm0.53 &   7.32\plm0.48  &    7.53\plm0.49  &   6.38\plm0.42   \\ 
58830.8669 &  20.28\plm1.14 &  16.38\plm0.76 &   10.46\plm0.59 &   7.67\plm0.50  &    7.67\plm0.50  &   6.94\plm0.45   \\
58837.5791 &    ---         &  10.43\plm0.58 &    7.44\plm0.49 &   7.39\plm0.48  &     ---          &   7.54\plm0.49  \\  
58851.0457 &  23.50\plm1.10 &  18.81\plm0.88 &   15.13\plm0.70 &  12.05\plm0.79  &   13.08\plm0.86  &  12.98\plm0.86   \\ 
58858.2817 &  24.38\plm1.14 &  19.16\plm0.89 &   15.84\plm0.74 &  14.22\plm0.94  &   14.48\plm0.95  &  15.04\plm0.99   \\
59312.1285 &  19.54\plm1.29 &  14.14\plm0.79 &    8.62\plm0.57 &   5.26\plm0.40  &    6.26\plm0.47  &   4.98\plm0.37   \\
59313.2613 &  19.37\plm1.09 &  15.08\plm0.70 &    8.08\plm0.45 &   5.98\plm0.39  &    5.40\plm0.41  &   4.62\plm0.35   \\
59758.5143 &  24.83\plm1.16 &  20.26\plm0.76 &   17.86\plm1.01 &  15.89\plm0.90  &   17.41\plm1.15  &  17.60\plm1.17\\
59855.6494 &  21.04\plm1.19 &  16.54\plm0.78 &   10.96\plm0.62 &   8.26\plm0.63  &    7.67\plm0.58  &   7.07\plm0.54\\
59859.3600 &  20.85\plm1.38 &  15.51\plm0.88 &   11.06\plm0.73 &   7.89\plm0.60  &    6.44\plm0.55  &   6.16\plm0.47\\
59918.0988 &  19.37\plm0.91 &  15.65\plm0.58 &    9.64\plm0.45 &   6.50\plm0.37  &    6.09\plm0.40  &   5.27\plm0.40\\
60042.2293 &  29.05\plm1.36 &  22.42\plm0.84 &   15.41\plm0.57 &  12.28\plm0.46  &   12.16\plm0.60  &  12.87\plm0.48\\
60046.5968 &  27.74\plm1.04 &  22.62\plm0.84 &   14.19\plm0.53 &  11.83\plm0.44  &   11.50\plm0.54  &  11.00\plm0.41\\
60047.2907 &  26.74\plm0.99 &  22.01\plm0.82 &   14.58\plm0.68 &  11.72\plm0.55  &   11.83\plm0.67  &  11.31\plm0.42\\
60051.1603 &  25.77\plm0.96 &  21.81\plm0.81 &   13.80\plm0.52 &  11.40\plm0.43  &   11.40\plm0.43  &  11.21\plm0.42\\
60052.3498 &  26.49\plm0.99 &  21.81\plm0.81 &   13.80\plm0.52 &  11.09\plm0.41  &   10.99\plm0.52  &  10.70\plm0.40\\
60063.6655 &  25.30\plm0.94 &  21.21\plm0.79 &   11.58\plm0.54 &   7.82\plm0.37  &    7.13\plm0.40  &   6.16\plm0.29\\
60770.7224 &  23.94\plm0.67 &  20.07\plm0.75 &   10.66\plm0.40 &   7.82\plm0.29  &    6.74\plm0.32  &   6.82\plm0.25\\
60772.1852 &  24.84\plm0.69 &  20.26\plm0.56 &   11.27\plm0.42 &   8.19\plm0.31  &    7.53\plm0.35  &   7.01\plm0.26\\
60773.2583 &  26.01\plm1.22 &  18.82\plm0.88 &   12.02\plm0.68 &   8.26\plm0.47  &    8.18\plm0.62  &   7.68\plm0.36\\
60834.9438 &  28.26\plm1.05 &  26.22\plm0.98 &   19.95\plm0.74 &  19.10\plm0.71  &   19.63\plm0.90  &  20.39\plm0.76\\
60846.8062 &  29.32\plm1.09 &  25.50\plm0.95 &   18.70\plm0.70 &  17.26\plm0.64  &   15.73\plm0.74  &  17.12\plm0.64\\
60860.2544 &  26.49\plm0.99 &  22.62\plm0.84 &   15.27\plm0.57 &  12.39\plm0.58  &   11.83\plm0.78  &  10.90\plm0.41\\
60874.9277 &  26.01\plm0.97 &  24.35\plm0.91 &   15.41\plm0.58 &  14.22\plm0.67  &   13.45\plm0.76  &  13.11\plm0.49\\
\noalign{\smallskip}
\hline                                                                          
                                     
\noalign{\smallskip}
\label{swiftmag}
\end{tabular}
\end{table}

\end{appendix}

\end{document}